\documentclass[twocolumn,aps,pra,showpacs,amsmath,amssymb,superscriptaddress]{revtex4-2}

\usepackage{comment}
\usepackage{graphicx} 
\usepackage{dcolumn}
\usepackage{booktabs}

\usepackage{csquotes}
\usepackage[colorlinks=true,linkcolor = {blue}, citecolor = {blue}, urlcolor = {blue}]{hyperref}
\usepackage{xcolor}
\usepackage[export]{adjustbox}
\usepackage{braket} 
\DeclareMathOperator{\Tr}{Tr} 

\newcommand{\victor}[1]{{\color{blue} {\bf V:} #1}}
\newcommand{\shah}[1]{{\color{red} {\bf S:} #1}}

\usepackage{ulem}

\usepackage{mathbbol}

\usepackage{orcidlink}

\begin{document}
\title{Robust Quantum Sensing with Multiparameter Decorrelation}

\author{Shah Saad Alam\orcidlink{0000-0001-6541-662X}}
\thanks{Authors S. S. A. and V. E. C. contributed equally}
\affiliation{JILA and Department of Physics, University of Colorado, 440 UCB, Boulder, CO 80309, USA}
\author{Victor E. Colussi\orcidlink{0000-0002-0972-6276}}
\thanks{Authors S. S. A. and V. E. C. contributed equally}
\affiliation{Infleqtion, Inc., 3030 Sterling Circle, Boulder, CO 80301, USA}
\author{John Drew Wilson\orcidlink{0000-0001-6334-2460}}
\affiliation{JILA and Department of Physics, University of Colorado, 440 UCB, Boulder, CO 80309, USA}
\author{Jarrod T. Reilly\orcidlink{0000-0001-5410-089X}}
\affiliation{JILA and Department of Physics, University of Colorado, 440 UCB, Boulder, CO 80309, USA}
\author{Michael A. Perlin\orcidlink{0000-0002-9316-1596}}
\affiliation{Infleqtion, Inc.,141 West Jackson Blvd
Suite 1875
Chicago, IL 60604, USA}
\author{Murray J. Holland\orcidlink{0000-0002-3778-1352}}
\affiliation{JILA and Department of Physics, University of Colorado, 440 UCB, Boulder, CO 80309, USA}

\date{\today}

\begin{abstract}
The performance of a quantum sensor is fundamentally limited by noise.
This noise is particularly damaging when it becomes correlated with the readout of a target signal, caused by fluctuations of the sensor's operating parameters.
These uncertainties limit sensitivity in a way that can be understood with multiparameter estimation theory.
We develop a new approach, adaptable to any quantum platform, for designing robust sensing protocols that leverages multiparameter estimation theory and machine learning to decorrelate a target signal from fluctuating off-target (``nuisance'') parameters.
Central to our approach is the identification of information-theoretic goals that guide a machine learning agent through an otherwise intractably large space of potential sensing protocols. 
As an illustrative example, we apply our approach to a reconfigurable optical lattice to design an accelerometer whose sensitivity is decorrelated from lattice depth noise. We demonstrate the effect of decorrelation on outcomes and Bayesian inferencing through statistical analysis in parameter space, and discuss implications for future applications in quantum metrology and computing.
\end{abstract}

\maketitle

\section{Introduction}\label{sec:intro}

Quantum sensing is a promising near-term application of quantum devices, with the potential to leverage the properties of quantum mechanics to revolutionize physical measurements \cite{PhysRevLett.131.150802,Holland1993HeisenbergLimit,RevModPhys.89.035002}.
This has been demonstrated in extensive applications such as atomic clocks~\cite{Jun_Zoller,Bothwell2022Redshift}, inertial sensors~\cite{Qvarfort2018Gravimetry,ReillyPDD,Richardson2020Optomech,Templier2022VectorAccel}, gravitational wave detectors~\cite{Aasi2013LIGO,Abbott2016LIGO,Abbott20162LIGO,Tse2019LIGO}, and biological tissue imaging devices~\cite{Taylor2016Bio}.
Quantum sensors also enable searches for new fundamental physics through, for example, precision tests of the equivalence principle and measurements of variations in fundamental constants \cite{yin2022experiments,schnabel2010quantum} to identify signatures of physics beyond the standard model~\cite{cairncross2017precision,roussy2023improved, safronova2018search}.
Common amongst these applications is the goal of improving upon the fundamental limitations posed by susceptibilities to classical and quantum noise (c.f.~\cite{PhysRevLett.124.060402,PhysRevA.98.030303,RevModPhys.90.035005, demkowicz2012elusive,DEMKOWICZDOBRZANSKI2015345,PhysRevLett.116.053601,zaheer2023quantum}).

\begin{figure}
    \centering
    \includegraphics[width=\linewidth]{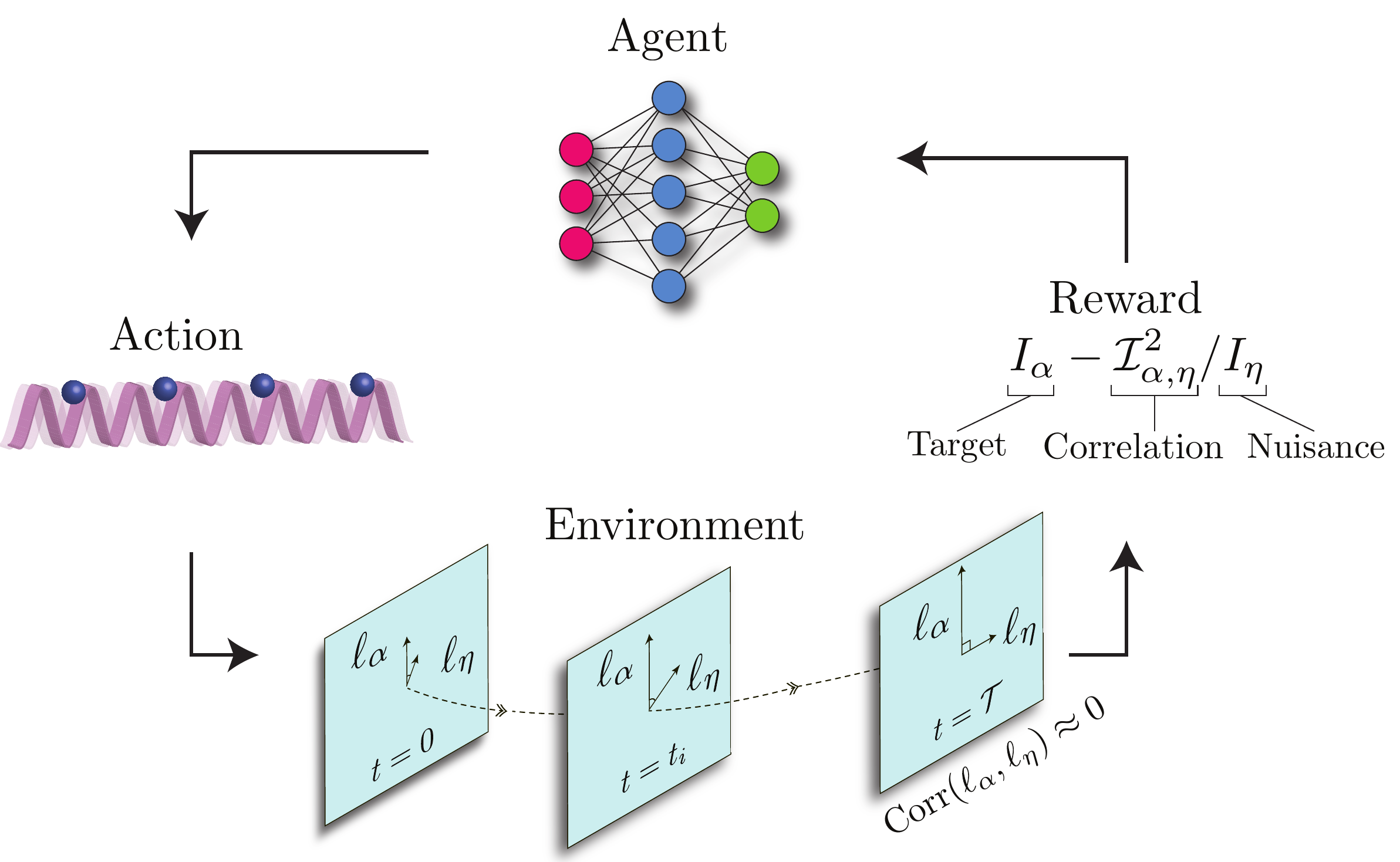}
    \caption{Machine learning framework for generating decorrelating sensing protocols (DSP's) using reinforcement learning in the multiparameter setting.  The agent first takes an action which generates a transition in the environment resulting in a reward.  The reward is formulated in terms of the score functions $\ell$ that determine the classical Fisher information matrix [see Eq.~\eqref{eq:scorecfim}].  This reward takes into account correlations between target and nuisance parameters in determining the overall sensitivity and decorrelating effect of the control function.  With each iteration, the reward is used to update the neural network and inform the decision making of the agent.} 
    \label{fig:conceptual}
\end{figure}

Past works have proposed many methods to
mitigate the effects of external noise.
These methods range from quantum error correction~\cite{dur2014improved,zhou2018achieving}, to classical compensation in lasers~\cite{siegman1986lasers,Genoni2011PhaseDiffusion}, to modifying the system dynamics through Hamiltonian steering methods~\cite{pang2017optimal}, and dynamical decoupling \cite{PhysRevLett.82.2417,souza2012robust,Jager2023measuring}.
More recently, programmable quantum sensors have been used to develop quantum metrology protocols to simultaneously resolve several parameters, such as the vector components of a field \cite{kaubruegger2019variational,marciniak2022optimal,PRXQuantum.4.020333, vec_acc_sci_advances}.
In this work, we design sensing protocols that minimize the classical correlations between off-target parameter uncertainties and the readout signal of interest.
Consequently, the adverse effects of both quantum and classical noise are reduced by ensuring that they
become statistically independent of the estimator for the target signal.

To engineer ``decorrelated" quantum sensing protocols (DSP's) that statistically isolate a target signal from other parameters, we task a machine learning agent with suitable information-theoretic goals \cite{Suzuki_2020}, as illustrated in Fig.~\ref{fig:conceptual}.
However, these protocols and rewards would work with any similar optimization protocol.
In the single-parameter setting, the study of sensing protocols from an information-theoretic perspective has already yielded new insights about how traditional paradigms \cite{PhysRevLett.124.060402,chih2022train,PhysRevA.98.023629,PhysRevLett.131.150802, robust_qoc} can fail to reach the full metrological potential available on a device.  
In particular, machine learning was used in Refs.~\cite{PhysRevLett.124.060402,chih2022train} to 
find schemes that maximized sensitivity using an end-to-end design that is unconstrained by traditional sensing paradigms, which involve distinct state-preparation and interrogation steps.

In this work, we unify these paradigms to propose a robust, decorrelating method that is agnostic to the choice of quantum device and can be applied to any multiparameter sensing task.
As an illustrative example, we consider an inertial sensor based on cold atom interferometry in an optical lattice, and use machine learning to make it robust to noise.
We design inertial DSPs to perform accelerometry in a setting of incomplete information about the optical lattice depth, for example due to online laser power noise, shot-to-shot variations in laser settings, or other (possibly unknown) effects.

\section{Background}\label{sec:mpest}
\subsection{Multiparameter estimation theory}\label{sec:mpecfim}
In this section we discuss multiparameter estimation theory, and establish the formalism used in the remainder of this work.
The goal of multiparameter estimation is to estimate the values of parameters $\mathbf{b} = (b^1,b^2,\cdots)$ that are encoded into a quantum state (density matrix) $\hat\rho(\mathbf{b})$.
In this work, we assume that the encoding occurs through time evolution via the equation of motion $\partial\hat\rho/ \partial t = \mathcal{L}_{\mathbf{b}}\hat\rho$, which takes an initial state $\hat\rho_0 \to \hat\rho(\mathbf{b})$.
The time evolution operator $\mathcal{L}_{\mathbf{b}}$ and final state $\hat\rho(\mathbf{b})$ implicitly depend on additional control parameters $\mathbf{c}$ that parameterize a sensing protocol.
Designing a sensing protocol thus entails choosing control parameters $\mathbf{c}$---such as a (discrete) choice of gates, or a (continuous) choice of a laser phase profile---that optimize the inference of the estimation parameters $\mathbf{b}$ from measurements of the state $\hat\rho(\mathbf{b})$.

Measurements of a quantum system are performed with respect to a basis, typically represented by a set of projectors $\mathbb{\Pi}=\{\hat{\Pi}_1,\hat{\Pi}_2,...\}$ \footnote{In fact, our results are unchanged if we replace the projector-valued measure (PVM) $\mathbb{\Pi}$ by a more general positive operator-valued measure (POVM).}.
Each projector is associated with a single measurement outcome, and the probability of obtaining outcome $j$ when measuring the state $\hat\rho(\textbf{b})$ is given by the likelihood function $P(j|\textbf{b}) = \Tr(\hat\rho(\mathbf{b})\Pi_j)$.
For a given prior $P(\textbf{b})$ that describes the \textit{a priori} likelihood (probability distribution) for different values of $\mathbf{b}$, Bayes' theorem \footnote{In this work we take a Bayesian approach to parameter estimation.
However, we emphasize that in the case of a fixed parameterization and flat prior, Bayesian and frequentist approaches are equivalent \cite{li2018frequentist}.} prescribes how measurement outcomes should be used to update the likelihood distribution for $\textbf{b}$.
For a sequence of statistically independent measurement outcomes $M = (m_1,m_2,...)$,
the conditional likelihood distribution for $\textbf{b}$ becomes
\begin{equation}
    P({\mathbf{b}}|M) =
    \frac{P(M|{\mathbf{b}})}{P(M)} P(\mathbf{b})=
    \left[\prod_{m\in M} \frac{P(m|{\mathbf{b}})} {P(m)} \right] P(\mathbf{b}),
\label{eq:Bayesian_updating_first}
\end{equation}
where $P(M|\mathbf{b})=\prod_{m\in M} P(m|\mathbf{b})$ is the joint likelihood function for outcomes $M$, and $P(M) =\prod_{m\in M} P(m)$ with $P(m)=\sum_{\bf b} P(m|{\bf b})$ is the probability of finding the measurement record $M$ for any possible value of $\mathbf{b}$.
Given a flat (uniform) prior $P({\bf b})$, the peak of the Bayesian posterior $P({\mathbf{b}}|M)$ coincides with the maximum likelihood estimator of the parameters:
\begin{equation}
    \hat{\mathbf{b}} = \mathrm{argmax}_{\mathbf{b}} \log P(M|{\mathbf{b}}).
    \label{eq:MLE}
\end{equation}

The ultimate sensitivity of a quantum sensor is decided by the sensitivity of each possible measurement outcome to a small change in a parameter.
This sensitivity is given by the score function
\begin{equation}
\ell_\mu (j|{\bf b}) \equiv \partial_\mu \log P(j|\mathbf{b}),
\end{equation}
where $\partial_{\mu} \equiv \partial/\partial b^\mu$.
The score function is used in the definition of the classical Fisher information matrix (CFIM) [see Fig.~\ref{fig:conceptual}]:
\begin{equation}\label{eq:scorecfim}
\begin{aligned}
& \mathcal{I}_{\mu\nu}(\mathbb{\Pi}|\textbf{b}) \equiv \mathbb{E}( \ell_\mu \, \ell_\nu |\textbf{b}) \\
&= \sum_{\Pi_j\in \mathbb{\Pi}} P(j|\mathbf{b}) \left[ \partial_{\mu} \log P(j
    |{\mathbf{b}}) \right] \left[ \partial_{\nu} \log P(j
    |{\mathbf{b}}) \right],
\end{aligned}
\end{equation}
where $\mathbb{E}(A|{\bf b}) = \sum_{\Pi_j\in \mathbb{\Pi}} P(j|\mathbf{b}) A (j|{\bf b})$ is the classical expectation value of $A$.
The CFIM quantifies the sensitivity of the probability distribution over measurement outcomes to small variations in the estimation parameters. 

\subsection{Optical-lattice interferometry}\label{sec:physmod}
We now consider multiparameter estimation in the context of accelerometry with an optical lattice interfereometer, which is a reconfigurable inertial sensing platform recently demonstrated experimentally in Ref.~\cite{ledesma2023machinedesigned}.  In this system, an ultracold atomic gas is loaded into the ground state of a one-dimensional optical lattice, described by the time-dependent Hamiltonian
\begin{align}
&\hat{H}(t) = \hat{H}_{\text{lat}}(t) + \hat{H}_{\text{acc}}(t),\label{eq:hamiltonian}\\ 
&\hat{H}_{\text{lat}}(t) = \frac{\hat{p}^2}{2m} - \frac{V_L}{2}\cos\left[2k_L \hat{x} + \phi(t)\right],\label{eq:hamiltonian_no_a} \\ 
&\hat{H}_{\text{acc}}(t) = ma\hat{x}, \label{eq:hamiltonian_a}
\end{align}
where $\hat{p}$ and $\hat{x}$ are the canonical one-body momentum and position operators, $m$ is the atomic mass, $V_L$ is the lattice depth, $k_L=2\pi/\lambda_L$ and $\lambda_L$ are the wavenumber and wavelength of the lattice light, respectively, $\phi(t)$ is a time-dependent phase called the ``control function'', and $a$ is the component of acceleration applied along the lattice axis \footnote{For simplicity, we do not consider the impact transverse components of the acceleration \cite{ledesma2023machinedesigned} in this work.}.
For concreteness, we consider an atomic gas of ${}^{87}\mathrm{Rb}$ atoms with mass $m\approx 86.9$ amu trapped in an optical lattice with wavelength $\lambda_L = 1064$ nm and a lattice depth of $V_L = 10E_R$ where $E_R= \hbar^2k_L^2/2m$ is the lattice recoil energy of the atoms.
The control function $\phi(t)$ is implemented with a time-dependent phase difference between the two counter-propagating beams that form the lattice. 

When the lattice is ``shaken'' with a time-varying control function $\phi(t)$, momentum and energy are exchanged via Bragg scattering between the lattice light and the atoms.  This transfer occurs in multiples of $2\hbar k_L$, which can be seen by expanding $\cos\left[2k_L \hat{x} + \phi(t)\right]\ket{p} \propto ( \ket{p+2 \hbar k_L} e^{i\phi(t)} + \ket{p-2 \hbar k_L} e^{-i\phi(t)} )$
\cite{PhysRevA.64.023604,PhysRevA.64.063613}.
Because the shaking is a global (translation-invariant) modulation of the lattice, the time-dependent control function does not alter the mean quasimomentum of the atoms (c.f.~\cite{PRXQuantum.2.040303}), and instead drives transitions between Bloch states.
Here, the energetically lowest (``insulating'' bands) Bloch states are weakly dispersive and not effective for bulk transport unlike the higher-lying (``conduction'' bands) states.  The flexibility of the control function allows multiple sensing tasks to be carried out on the same device, making the platform reconfigurable.
At the end of the shaking sequence, the occupations of different momentum states are recorded and can, neglecting interactions, taken to be a measurement record $M$ of size $N_{\mathrm{atoms}}$ independent single atom measurements where $N_{\mathrm{atoms}}$ is the total number of atoms in the wavepacket.
The measurement record, in turn, is used for Bayesian inferencing of the acceleration $a$ and lattice depth $V_L$ \footnote{Data and code for Sec.~\ref{sec:stats} are available for reference at Ref.~\cite{alam2024lattice_evolution}}.

\subsection{Previous works}\label{sec:refsys}

Accelerometry using optical lattice interferometers has been performed within the Mach-Zehnder configuration ~\cite{PhysRevResearch.3.033279, ledesma2023machinedesigned, 10015539, 9867736, 10156455, PhysRevLett.120.263201, Weidner_2018}.  Here, machine learning techniques were used to obtain control functions that produce beam-splitting and mirroring components.  The reconfigurability of the platform however allows for other, non-traditional protocols and paradigms to be considered, which may lead to improved sensitivity.

As discussed in Ref.~\cite{PhysRevA.98.023629}, for conventional three-pulse accelerometers and gravimeters operating in free space \cite{BORDE198910,PhysRevLett.67.181}, the Mach-Zehnder inteferometer (MZI) result $I_\mathrm{MZI}(p_0) = (2p_0 T^2/\hbar)^2$ for splitting $p_0$, free evolution time $T = \mathcal{T}/2$, and total operating time $\mathcal{T}$, does not produce the optimal sensitivity from the standpoint of the Fisher information \cite{PhysRevA.98.023629}.  In that work, it was found that removal of the mirroring pulse lead instead to Ramsey interferometry (RI), boosting the sensitivity as $I_\mathrm{RI}(p_0) = 4I_\mathrm{MZI}(p_0)$.  However, this configuration requires a momentum rather than (conventional) spin population measurement, with the latter necessitating a spatial overlap in the output for a high contrast signal.  The relative compactness of optical-lattice interferometer allows for resolution of momentum-space occupations directly via time-of-flight measurements or in-trap via Bragg or Doppler spectroscopy (c.f.~\cite{ernst2010probing,PhysRevA.91.023626,PhysRevLett.82.4569, PhysRevLett.107.175302,PhysRevA.61.063608}) such that optimal sensitivity in the Ramsey configuration is in principle achievable.  The demonstration of {\it in-situ} detection methods on this device remains the subject of future study.

Nevertheless, a roadblock in achieving the optimal sensitivity of the Ramsey configuration on an optical lattice interferometer is the long component times compared to total operating time.   Additionally, the readout is constrained by the fixed basis of the experiment, and the management of these competing constraints is a task well-suited for machine learning techniques.  In this paper, we pursue the problem of achieving the Ramsey bound while decorrelating fluctuating off-target parameters from spoiling the sensitivity of the device to inertial signals.  In the following section, we describe a novel approach to decorrelation within the framework of multiparameter estimation theory described in Sec.~\ref{sec:mpecfim}.

\section{Decorrelated Sensing Protocols}\label{sec:decsens}

Having provided the requisite background concepts in Sec.~\ref{sec:mpest}, we now outline the main finding of this work: that the CFIM can be cast as an information-theoretic goal suitable for input into machine learning techniques to produce decorrelated sensing protocols (DSP's).  In the present case, these protocols consist of control functions $\phi(t)$ that dictate the dynamical modulation of the optical lattice for performing accelerometry.  

Recall that the CFIM relates the response of the possible measurement outcomes to small variations in the parameters.  Explicitly, the posterior can be expanded as
\begin{equation}\label{eq:postexp}
   \log P({\bf b}+ d{\bf b}|M) - \log P({\bf b}|M) = -\frac{N}{2} \ \mathcal{I}_{\mu\nu} \ db^\mu db^\nu,
\end{equation}
where $N=|M|$ is the length of the measurement record, we implicitly sum over repeated indices, and for brevity we henceforth suppress the explicit dependence of the CFIM $\mathcal{I}$ on measurement basis $\mathbb{\Pi}$ and estimation parameters $\mathbf{b}$.
Eq.~\eqref{eq:postexp} illuminates the geometrical role of the CFIM as a metric on smooth statistical manifolds \cite{cover1999elements}.
Physically, Eq.~\eqref{eq:postexp} reveals that the resolvability of parameter values, and hence the ultimate sensitivity of the device, is related to the informational distance between the two posteriors, and hence by the CFIM which determines distances across the informational landscape. 

Correlations between parameters are quantified by the covariance matrix $\mathrm{cov}(\mathbf{b}|M)$, whose value at row $\mu$ and column $\nu$ is the covariance between the estimates of parameters $b^\mu$ and $b^\nu$, found from Eq.~\eqref{eq:MLE}.
The diagonal entry with $\nu=\mu$ is just the variance of the estimate of $b^\mu$.
The asymptotic lower bound for this uncertainty is given by the (multiparameter) Cram\'er-Rao bound~\cite{Liu_2020,holevo2011probabilistic,doi:10.1080/23746149.2016.1230476}, which can be expressed by the inequality
\begin{equation}\label{eq:ccrb}
\mathrm{cov}({\mathbf{b}}|M) \geq \frac{1}{N}\mathcal{I}^{-1} \geq \frac{1}{N} \mathcal{F}^{-1},
\end{equation}
where $\mathcal{I}^{-1}$ is the inverse of the CFIM, and $A\geq B$ means the matrix $A-B$ is positive semi-definite (has non-negative eigenvalues)
\footnote{In the case of zero diagonal entries in the CFIM/QFIM, these are omitted when calculating the Cramer-Rao Bound. In practical systems, however, it'll be rare to have zero Fisher information}. The minimal bound in Eq.~\ref{eq:ccrb} is achieved by $\mathcal{F}$, the quantum Fisher Information matrix (QFIM), which represents the matrix upper bound of $\mathcal{I}$.
The QFIM is constructed by maximizing each diagonal element $\mathcal{I}_{\mu \mu}$ over \textit{all} possible sets of measurements.
In practice, this means replacing the score functions for every element of the matrix in Eq.~\eqref{eq:scorecfim} with parameter derivatives of the quantum state.
For a pure state, this takes the form ~\cite{Liu_2020}
\begin{equation}\label{eq:qfimoffdiag}
\mathcal{F}_{\mu\nu} = 4\mathrm{Re}\left[\langle \partial_\mu \psi|\partial_\nu \psi\rangle - \langle \partial_\mu \psi|\psi\rangle\langle \psi |\partial_\nu\psi\rangle\right].
\end{equation}


The presence of parameter correlations leads to finite off-diagonal values of the CFIM, producing a ``tilt" of the Bayesian posterior, which we quantify by the dimensionless correlation
\begin{equation}
\begin{aligned} \label{eq:correlation_ratio}
\mathrm{Corr}(\ell_\mu,\ell_\nu) \equiv& \frac{\mathcal{I}_{\mu\nu}}{\sqrt{I_{\mu}} \sqrt{I_{\nu}}},
\end{aligned}
\end{equation}
where $I_\mu = \mathcal{I}_{\mu\mu}$, for brevity.
This quantity diagnoses how correlated the parameter score functions for $b^\mu$ and $b^\nu$ are (as indicated in Fig.~\ref{fig:conceptual}), with $+1$ being perfectly correlated, $0$ being decorrelated (statistically independent), and $-1$ being anti-correlated.  We note that the connection between Eqs.~\eqref{eq:postexp} and \eqref{eq:correlation_ratio} becomes apparent in the normalized parameterization $\tilde{\textbf{b}}^\mu \equiv b^\mu/\sqrt{I_\mu}$, where the dimensionless correlation enters Eq.~\eqref{eq:postexp} in place of $\mathcal{I}_{\mu\nu}$.

\subsection{Target and Nuisance Parameters}
We now develop further language specific to the task of parameter decorrelation.  The operation of a quantum sensor will, in general, depend on multiple parameters but only some of these will be of interest.
The parameters of interest are called \textit{target} parameters and are denoted by the ordered list $\boldsymbol{\alpha}=(\alpha^1,\alpha^2,...)$, while the other parameters are called \textit{nuisance} parameters and are denoted by the ordered list $\boldsymbol{\eta}=(\eta^1,\eta^2,...)$, so the total ordered list of parameters is $\mathbf{b} = (\alpha^1,\alpha^2,...,\eta^1,\eta^2,...)$.
Our ability to measure the target parameters depends on our degree of knowledge and control over all parameters, some of which are changing or initially unknown, such as lattice fluctuations in lattice atomic clocks~\cite{takamoto2005optical,goban2018emergence}.

Even if nuisance parameters are removed via marginalization, the uncertainties in the nuisance parameters can feed into the uncertainty for the target parameters via correlations.
To see this, consider the general case of multiple target parameters for which we wish to marginalize out the dependence on the nuisance parameters.
To do so, we consider the block form of the CFIM\cite{Suzuki_2020},
\begin{equation}
\mathcal{I} = \begin{pmatrix}
\mathcal{A} & \mathcal{C}\\
\mathcal{C}^\text{T} & \mathcal{N}
\end{pmatrix},
\end{equation}
where $\mathcal{A}$ and $\mathcal{N}$ are the CFIM's for only $\boldsymbol{\alpha}$ and $\boldsymbol{\eta}$ respectively, and $\mathcal{C}$ is the cross terms, i.e. $\mathcal{C}_{\mu j}$ is the correlation between $\alpha^\mu$ and $\eta^j$.
By the block matrix inversion theorem, the inverse Fisher information that appears in the Cram\'er-Rao bound of Eq.~\eqref{eq:ccrb} is
\begin{equation} \label{eq:BlockInversion}
\left(\mathcal{I}^{-1} \right)^{\mu\nu} = \left( \mathcal{A}_{\mu\nu} - \mathcal{C}_{\mu i} \left( \mathcal{N}^{-1}\right)^{ij} \mathcal{C}_{j \nu} \right)^{-1},
\end{equation}
where Greek indices $\mu$, $\nu$ run over target parameters and Latin indices $i$, $j$ are summed over nuisance parameters\cite{Suzuki_2020}.

As an illustrative example, consider the case of just one target parameter $\alpha$ and nuisance parameter $\eta$, so $\mathbf{b} = (\alpha,\eta)$.
If $\eta$ were perfectly known, then it is straightforward to saturate the Cram\'er-Rao bound--although often difficult in real experimental contexts.
However, it is rarely the case that all other parameters are known.
In this case, even though we don't necessarily care about these other parameters, we have~\cite{bobrovsky1987CRBs, Suzuki_2020}
\begin{equation}
\mathrm{var}(\alpha) \geq \left[ N I_{\alpha} \left( 1 - \mathrm{Corr}(\ell_{\alpha},\ell_{\eta})^2 \right) \right]^{-1},
\end{equation}
where we've used the parameters as labels, rather than indices.  This demonstrates the penalizing effect of having incomplete information about additional parameters in a system, even when one is only concerned about a single parameter.

\begin{figure*}[t!]
	\centering
    \includegraphics[clip, width = \textwidth]{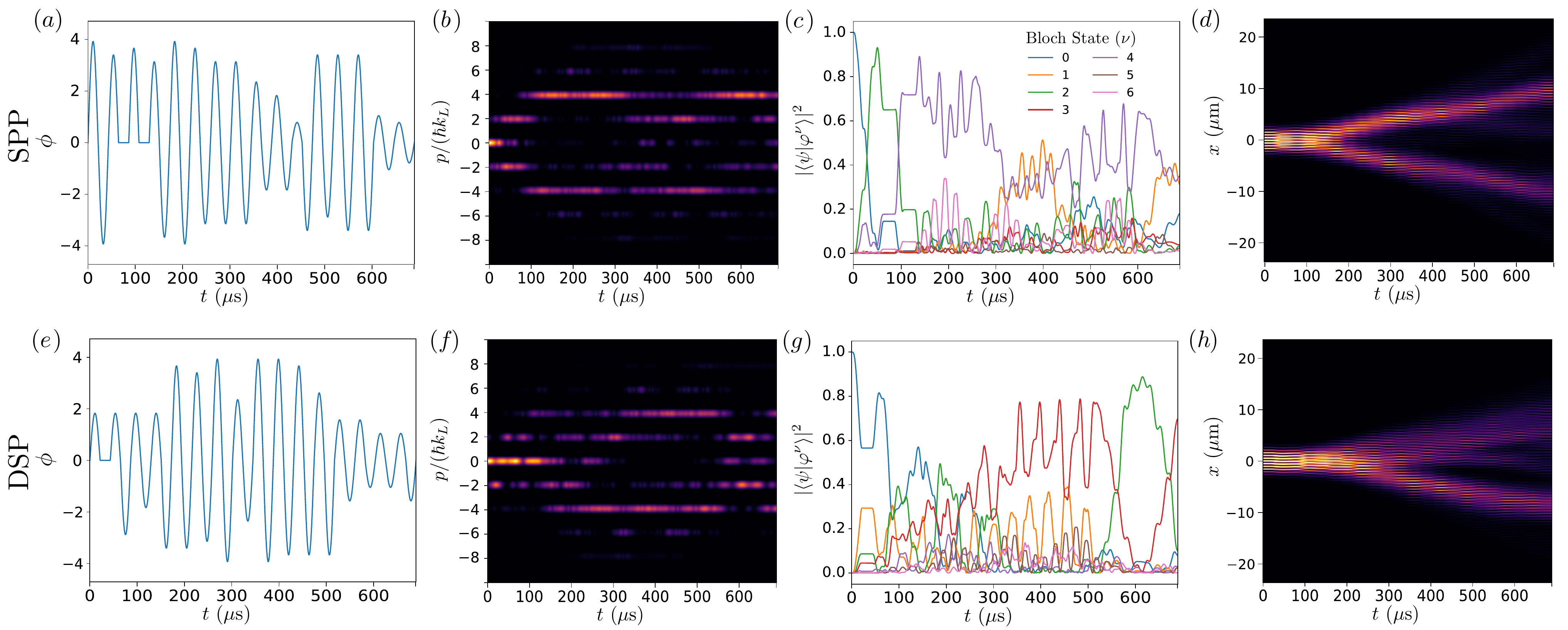}
    \caption{Accelerometry using the SPP (a-d) and DSP (e-h), including control functions (a,e), momentum space dynamics (b,f), Bloch state occupations (c,g), and position space dynamics (d,h).  Position space visualizations in this work are made by the convolution of $\psi_q(p,t)$ with a Gaussian of width $\sigma_p = 0.1\hbar k_L$. This illustrates the split trajectories, initialized as a Gaussian wave packet spanning roughly 10 lattice sites. 
    In practice (c.f.~\cite{ledesma2023machinedesigned}) the initial wave packet is delocalized over many times this number of sites such that trajectories are not resolvable.}		\label{fig:evo_a}
\end{figure*}

\begin{figure}[t!]
	\centering
    \includegraphics[clip, width = 1 \columnwidth]{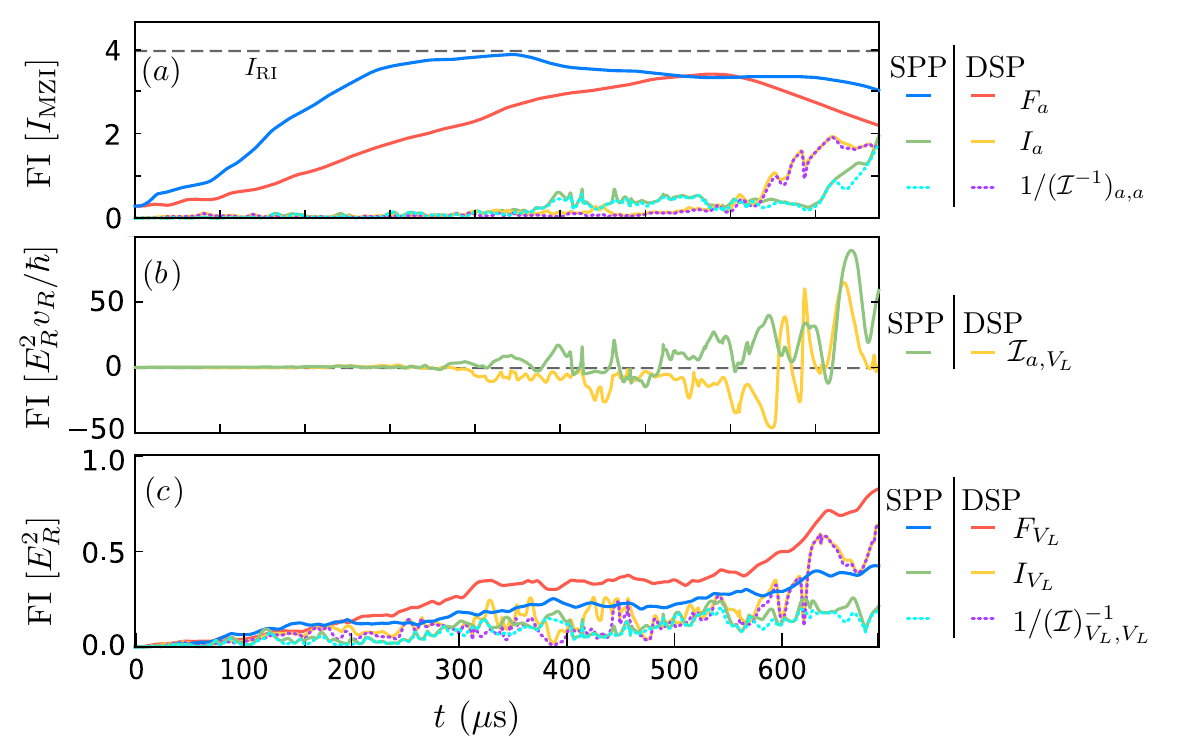}
     \caption{Relevant classical and quantum Fisher information (FI) elements for (a) acceleration, (b) covariance, and (c) lattice depth in the SPP and DSP cases.  Here, $v_R = \hbar k_L/m$ is the recoil velocity.  Panel (a) is in units of the Mach-Zehnder interferometer to illustrate advantage over traditional setups, and the dashed line indicates the optimal result of the Ramsey configuration discussed in Sec.~\ref{sec:refsys}.}		\label{fig:fi_a}
\end{figure}

The goal of a DSP is then to optimize sensitivity to a small change in one or many target parameters in $\boldsymbol{\alpha}$, while also being invariant to changes of any nuisance parameters in $\boldsymbol{\eta}$.
From Eq.~\eqref{eq:BlockInversion} we see that this is equivalent to maximizing some property of $\mathcal{A}$--such as its smallest eigenvalue, its trace, or its determinant depending on the specific use case~\cite{pukelsheim2006optimal}, while minimizing some property of $\mathcal{C} \mathcal{N}^{-1} \mathcal{C}^\text{T}$.  In this work, we choose to input elements of Eq.~\eqref{eq:BlockInversion} directly into a machine learning method for developing DSP's in the case of one target and one nuisance parameter [see Fig.~\ref{fig:conceptual}].  Further details on the machine learning method employed can be found in App.~\ref{app:rl}.  Finally, we comment that the formalism developed in this section for designing DSP's targets two common forms of nuisance parameter fluctuations:  miscalibration and shot-to-shot parameter drift such as frequency drift, laser phase noise in an atomic clock, or lattice noise in an optical lattice system~\cite{al2015noise}. 


\subsection{Accelerometry}\label{sec:mpqs}

In this section, we analyze the quantum dynamics that result from a DSP designed to perform accelerometry in the presence of fluctuations in the nuisance parameter $V_L$.  In this case, the appropriate element of Eq.~\eqref{eq:BlockInversion} to be used as design goal for developing DSP's for accelerometry using machine learning (see Fig.~\ref{fig:conceptual}) is
\begin{equation}\label{eq:designgoal}
\text{Design Goal = }\underset{\phi(t)}{\mathrm{max}}\left[1/(\mathcal{I}^{-1})_{a,a} = I_a - \frac{\mathcal{I}_{a,V_L}^2}{I_{V_L}}\right].
\end{equation}
As discussed in App.~\ref{app:rl}, we use reinforcement learning to search for control parameters $\mathbf{c}$ corresponding to a control function $\phi(t)$ that maximizes this goal~\footnote{We note that the formulation in Eq.~\eqref{eq:designgoal} and Eq.~\eqref{eq:BlockInversion} is written generally for application with a generic optimizer.}.
We judge the performance of the DSP based on sensitivity to the target parameter $\zeta_a = 1/(\mathcal{I}^{-1}_{a,a}I_{MZI})$ as well as reduction in the value of the dimensionless correlation $\mathrm{Corr}(\ell_a,\ell_{V_L})$ introduced in Eq.~\eqref{eq:correlation_ratio}.  To illustrate the effect of decorrelation, we compare results where perfect information about nuisance parameters exists, using the single-parameter protocol (SPP) shown in Fig.~\ref{fig:evo_a}(a), against results where we have no information about the nuisance parameters, using the DSP shown in Fig.~\ref{fig:evo_a}(e).  We expect the dimensionless correlation to take random values between $[-1,1]$ for the SPP, as the elements of Eq.~\eqref{eq:BlockInversion} supplied to the machine learning method contain information only about the target parameter.  Instead, we expect this ratio to instead be close to $0$ for a successful DSP.   Additionally, since we wish to compare against reference sensing protocols as well as previous works on optical lattice interferometry \cite{ledesma2023machinedesigned}(see Sec.~\ref{sec:refsys}), we reject control functions that generate occupations in excess of a few percent for  momentum states with magnitudes greater than $p_0 = 4\hbar k_L$.

Fig.~\ref{fig:evo_a} shows the time-dependent control functions learned by the reinforcement learning agent, as well as the resultant quantum state dynamics, for both the SPP [(a)-(d)] and DSP [(e)-(h)].  Despite differences in the control functions, within 200 $\mu$s the quantum state is excited from valence $(\nu = 0,1,2)$ to conduction bands $(\nu \geq 3)$ where bulk spatial transport becomes possible as shown by the split spatial trajectories in Fig.~\ref{fig:evo_a}~(d,h).  During this transport phase, the Bloch band occupation dynamics differ greatly between the protocols as shown in Fig.~\ref{fig:evo_a}~(c,g), with significant occupations of the lowest conduction bands, which are superpositions of the form $|p_0,\pm\rangle \equiv (|p_0\rangle \pm |-p_0\rangle)/\sqrt{2}$.
By $t\sim 550$ $\mu$s, the quantum state is prepared for measurement in the fixed basis of the experiment.  Here we see that in the SPP, population is transferred to the valence bands, whereas in the DSP a large oscillation occurs between valence and conduction bands, crucial for decorrelation of nuisance and target parameters.  


From Figs.~\ref{fig:evo_a}(d,h) we see that  a large spacetime area is enclosed without an obvious mirroring component, characteristic of the Ramsey configuration discussed in Sec.~\ref{sec:refsys}.  This leads in both cases to a sensitivity $\zeta_a\approx 2$, twice that of the conventional Mach-Zehnder configuration but half that of the Ramsey configuration.  Analysis of the Fisher information matrices (quantum and classical) shown in Fig.~\ref{fig:fi_a}, reveals in panel (a) that the optimal result $I_{RI}$ is approached in both cases where the QFIM is evaluated using Eq.~\ref{eq:qfimoffdiag}
However the final sensitivity is limited by the finite duration of the basis preparation phase.  Fig.~\ref{fig:fi_a}(b,c) reveals instead the decorrelating effect of the DSP ($\mathrm{Corr}(\ell_a,\ell_{V_L})\approx 0.004$) compared to the SPP ($\mathrm{Corr}(\ell_a,\ell_{V_L})\approx 0.3$).  

From Fig.~\ref{fig:fi_a}, we see that the DSP also yields more information about the lattice depth.  This is an additional strategy to further reduce the dimensionless correlation, i.e. by executing a sensing protocol that reduces our uncertainty in target and nuisance parameters alike, thus reducing the multiparameter estimation problem to that of a single parameter where an SPP becomes more appropriate.    
    

\section{Statistical Analysis}\label{sec:stats}

In this section, we analyze the utility of DSPs for performing accelerometry in the presence of fluctuations and uncertainty in the lattice depth.  In Sec.~\ref{sec:bayes}, we use Bayes' theorem (see Sec.~\ref{sec:mpecfim}) to estimate the target parameter $a$ from the final momentum distributions obtained by running a DSP over a range of accelerations and lattice depths.  In Sec.~\ref{sec:js}, we further quantify performance and establish the range of validity of the DSP through an analysis of the Jensen-Shannon divergences (JSD).  This analysis extends to the multiparameter domain approaches developed in Refs.~\cite{ledesma2023machinedesigned,chih2022train,PhysRevResearch.3.033279} to handle the complex, multi-modal output of an optical lattice interferometer.

\begin{figure}[t!]
    \centering
    \includegraphics[width=0.9\linewidth]{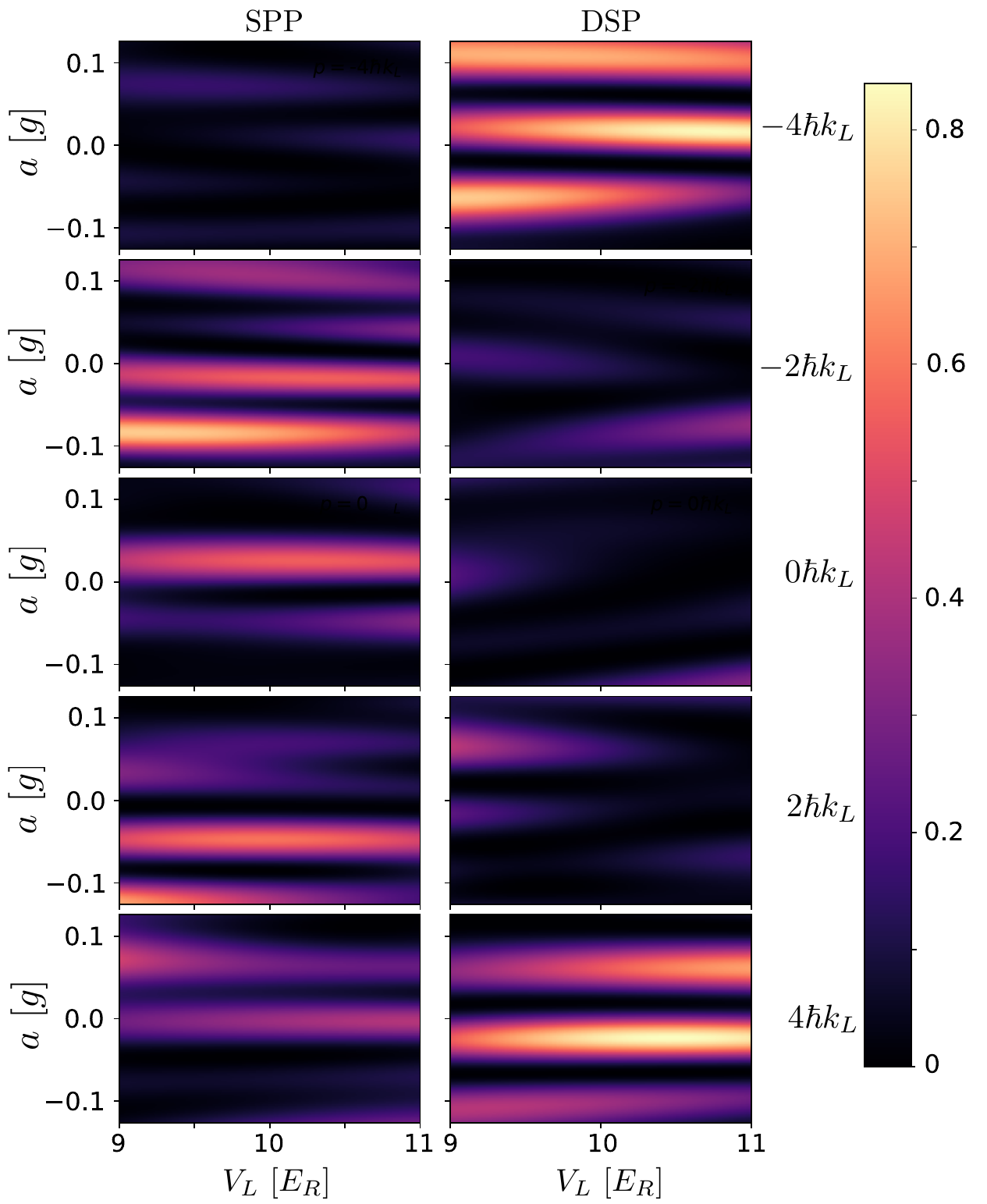}
    \caption{Final momentum probability distributions $P(p|a,V_L)$, resulting from the SPP (left column) and DSP (right column) presented in Fig.~\ref{fig:evo_a} evaluated over a range of $a$, $V_L$, and $p$.  Unlike a conventional Bragg or three-pulse interferometer that outputs a single interference signal, an optical lattice interferometer returns fringes for each momentum component obtained via time-of-flight measurements.  For clarity we display occupations up to momenta $\pm 4\hbar k_L$ and note that larger momenta can also be occupied particularly as the acceleration or lattice depth is varied significantly.  Although not shown, these occupations are accounted for in the analysis of this section.}
    \label{fig:2d_fringes}
\end{figure}

\subsection{Bayesian estimation}\label{sec:bayes}

\begin{figure}
    \centering
    \includegraphics[width=\linewidth]{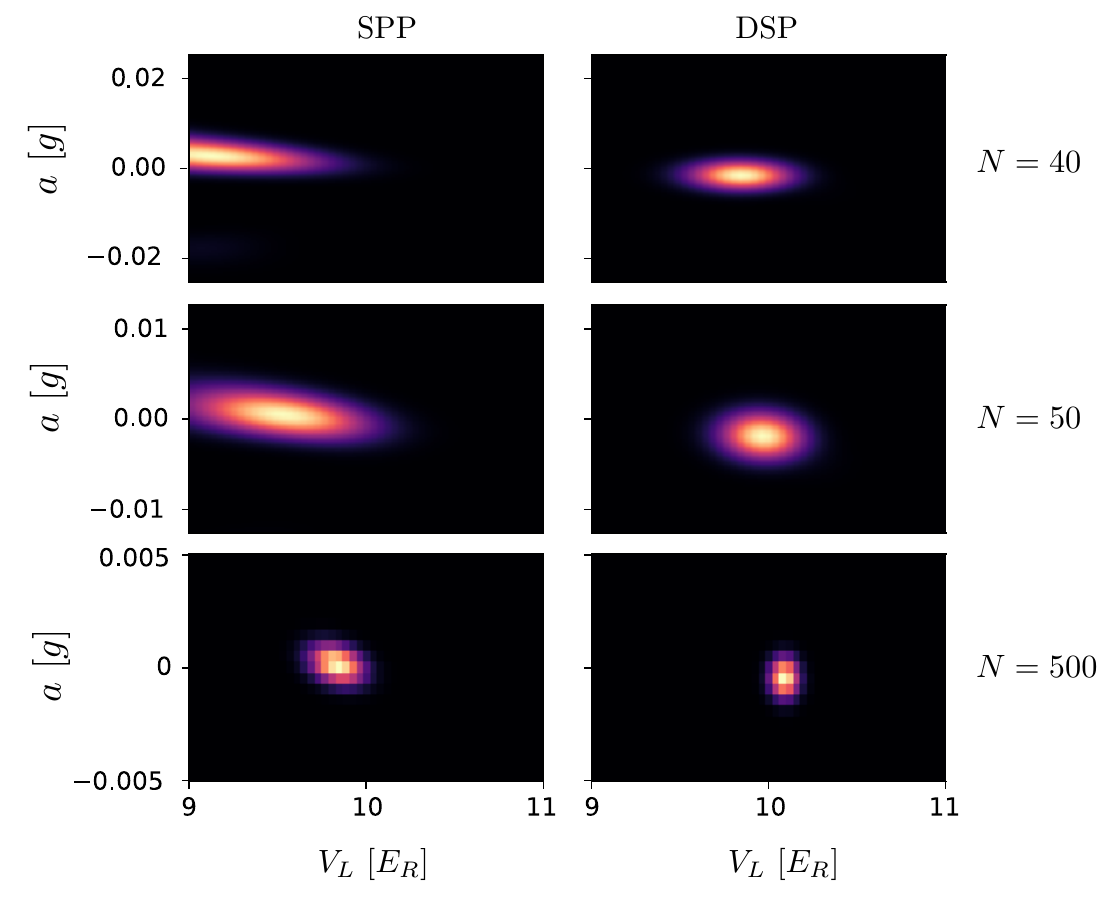}
   \caption{Bayesian updating in the multi-parameter estimation problem using the SPP (left column) and DSP (right column) as a function of measurement number $N$.  In each subplot, the posterior probability distribution (Eq.~\eqref{eq:Bayesian_updating_first}) is shown, which narrows on the actual parameter values $(a,V_L) = (0 g,10 E_R)$ after repeated measurement, allowing for parameter estimation via the maximum likelihood estimate (Eq.~\eqref{eq:MLE}).  For sufficiently large $N$, the uncertainty in the posteriors obeys the Cram\'er-Rao bound (Eq.~\eqref{eq:ccrb}) with the visible tilt of the SPP a consequence of the relatively large value of $\mathcal{I}_{a,V_L}$ for this sequence compared to the DSP.  As discussed in Sec.~\ref{sec:decsens} this tilt is a reflection also of the dimensionless correlation.  We note that each subplot has been normalized to provide visible contrast of the posteriors. 
   }
    \label{fig:2param_Bayesian_prob}
\end{figure}

We begin by performing Bayesian estimation (Eq.~\eqref{eq:Bayesian_updating_first}) on the final momentum distributions $P(p|a,V_L)$, which are shown in Fig.~\ref{fig:2d_fringes} for the SPP and DSP evaluated over the regions $-0.1\leq a/g \leq 0.1$ and $9\leq V_L/E_R\leq 11$.  From Eq.~\eqref{eq:postexp}, the rate of variation of the distributions in parameter space can be understood in terms of the CFIM, with the fast variation in $a$ indicative of protocols designed for performing accelerometry (rather than lattice depth estimation).  The multi-modal, multi-parameter nature of the distributions shown in Fig.~\ref{fig:2d_fringes}, however, complicates the statistical analysis and interpretation of the CFIM.  The results of multiparameter Bayesian updating using these distributions are shown in Fig.~\ref{fig:2param_Bayesian_prob}.  Here, the posterior distribution is shown for a measurement record $M$ assembled after taking $N$ stochastic samples of the distribution function $P(p|a,V_L)$ evaluated at the point $(a, V_L) = (0g, 10E_R)$, corresponding to the coordinates used for optimization of the DSP in Sec.~\ref{sec:mpqs}.

Again, the resulting posterior distributions resulting after $N$ measurements can be understood via Eq.~\eqref{eq:postexp}, where now the differences between the final CFIM elements for the SPP and DSP (see Fig.~\ref{fig:fi_a}) are visually evident.  Both protocols were designed to perform accelerometry, with the large sensitivity $\zeta_a$ reflected in the ``narrowness'' of the posterior over the range of accelerations considered.  Indeed, as additional measurements are performed, the standard deviation of the distribution approaches the standard quantum limit $\sim 1/\sqrt{N}$ according to the Cram\'er-Rao bound (Eq.~\eqref{eq:ccrb}) which determines the lower bounds on sensitivity in terms of the CFIM.  Here, the noticeable ``tilt'' of the SPP posterior is a reflection of the larger values of $\mathcal{I}_{a,V_L}$ and dimensionless correlation relative to the DSP as discussed in Sec.~\ref{sec:mpqs}.  We recall from that section that the dimensionless correlation differs by two orders of magnitude between SPP and DSP. As a consequence, Fig.~\ref{fig:2param_Bayesian_prob} provides a visually intuitive confirmation that the desired decorrelating effect is achieved by the DSP in the maximum likelihood estimation.  Furthermore, we comment that the reduction of uncertainty in target and nuisance parameters alike by DSP's leads to a narrower posterior along the $V_L$ axies for the DSP. 

A limitation of the Bayesian analysis shown in Fig.~\ref{fig:2param_Bayesian_prob} is that it only reveals performance of the accelerometer locally around the parameters of the simulated measurement record.  It is important, however, to also consider device operation using the DSP over a broader range of parameter space, which we consider in the following section.

\subsection{Jenson-Shannon Divergence}\label{sec:js}
Generally speaking, the precision of a quantum sensor is determined by how rapidly the measurement outcomes change with small variations $d{\bf b}$ of parameters. Consequently, if the outcomes $P(p|{\bf b})$ and $P(p|{\bf b} + d{\bf b})$ are too ``close'', the sensor may fail to distinguish different regions in parameter space during Bayesian estimation.  To quantify this statistical notion of closeness, we analyze the Jenson-Shannon divergence (JSD) which, for two probability distributions $P$ and $Q$, is given by 
\begin{equation}\label{eq:jsd}
    D_{JS}(P||Q)= \frac{1}{2} \left[S( P||M) + S( Q||M) \right ],
\end{equation} 
where $M=\frac{1}{2}(P+Q)$ is a mixture of $P$ and $Q$, and $S(A||B)=-\sum_p A(p) \log_2 \frac{B(p)}{A(p)}$ is the relative entropy (also known as the Kullback–Leibler divergence) of $A$ with respect to $B$. 
The JSD captures the shared information between $P$ and $Q$ given measurements from a mixture of the two. Intuitively, a small value of the JSD means the two probability distributions almost perfectly coincide, and therefore would be difficult to resolve via Bayesian estimation.  In this section, we extend the usage of the JSD for optical lattice interferometry from Ref.~\cite{ledesma2023machinedesigned} to performing accelerometry in the multiparameter domain, while emphasizing that the discussion applies more generally to characterizing any DSP where information about a (potentially) large space of target and nuisance parameters must be compressed and analyzed.


We first relate the JSD to the CFIM by expanding for small changes in parameters for the case of accelerometry considered in the present work (see App.~\ref{app:jsdexpand} for general expressions)
\begin{align}\label{eq:curvature}
D_{JS}(P(p|a,V_L)&||P(p|a+da,V_L+dV_L))\nonumber\\
&\approx
\frac{1}{8}\left[I_a da^2 + I_{V_L}dV_L^2 + 2\mathcal{I}_{a,V_L}da\ dV_L\right].
\end{align}
The CFIM determines the local curvature of the JSD, relating small parameter variations to displacements on the smooth statistical manifold~\cite{cover1999elements}, impacting Bayesian estimation (Eq.~\eqref{eq:postexp}), and the decorrelization effect of the protocol. Expanded instead using the normalized parameterization ${\bf \tilde{b}}$ discussed in Sec.~\ref{sec:mpqs}, the matrix element $\mathcal{I}_{a,V_L}$ is replaced by the dimensionless correlation, highlighting the universal relevance of this quantity.  

\begin{figure}[t!]
     \centering
     \includegraphics[width=\linewidth]{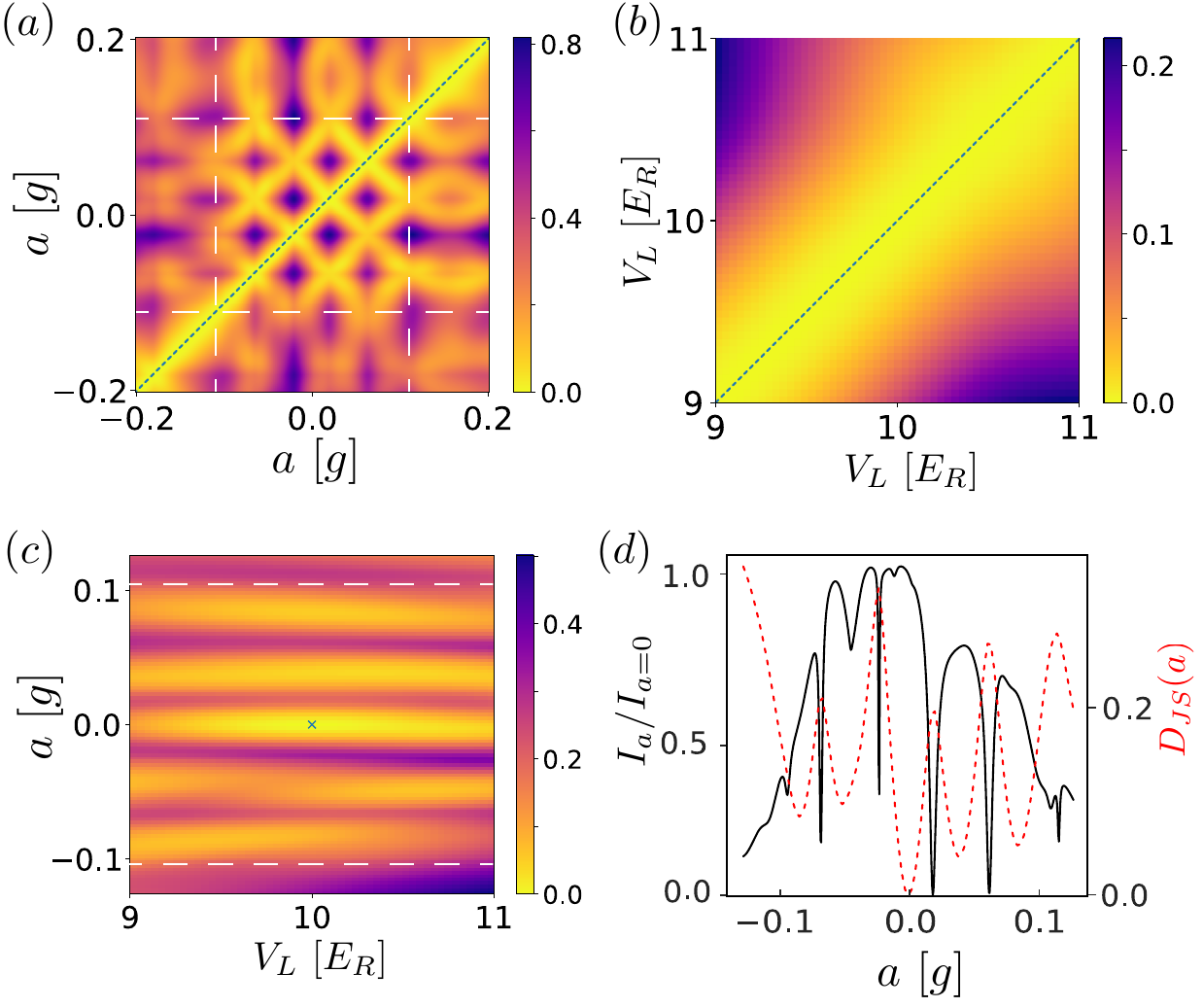}
        \caption{
    Visualization of the Jensen-Shannon divergence for the DSP as pairs of parameters are varied, including (a) fixed lattice depth and varying acceleration, (b) fixed acceleration and varying lattice depth, and (c) varying acceleration and lattice depth.  The non-varied parameters are held fixed at the $a/g = 0$ and $V_L/E_R = 10$ for consistency.  In panels (a) and (c), dashed white lines indicate regions where the phase repeats by $2\pi$, denoting the effective range of the sensor for performing accelerometry.  In (c), we plot the JSD around the training point for DSP (indicated with a cross)
    whose local curvature matches the final CFIM elements presented in Fig.~\ref{fig:fi_a}.  In panels (a) and (b), the blue dotted lines indicate the main diagonal where the JSD is vanishing due to perfect coincidence of the probability distributions. Panel (d) shows the $D_{JS}(P(p|a,V_L = 10E_R)||P(p,a'=0g,V_L = 10E_R)$ as a function of acceleration $a$ in dotted red lines, which is equivalent to a evaluating a horizontal slice of (a) at $a'=0$. The local $I_a$ along this slice is also shown as a black solid line. We see that the CFI envelope decays signficantly by $a/g= \pm 0.11$ and the JSD envelope increases along the same range, allowing an estimation of the sensor's effective range.}
\label{fig:Jenson-Shannon}
\end{figure}

Having established the local connection between JSD and CFIM, we now plot the JSD over finite domains of interest in parameter space for centered around the coordinates used for optimizing the DSP from Sec.~\ref{sec:mpqs}.  In principle, this requires a four-dimensional analysis however, we consider instead variations of pairs of parameters in order to illustrate our main findings.  Figure~\ref{fig:2d_fringes}(a) shows the JSD with lattice depth fixed at the optimized coordinate with varying accelerations.  The sensitivity of the protocol is reflected in the curvature surrounding the diagonal according to Eq.~\eqref{eq:curvature}) which depends directly on $I_a$.  Away from the diagonal, regions of nearly-vanishing JSD are also observed, indicating repetition of likelihood functions at different locations in parameter space.  The low JSD off-diagonals arise due to the spatial inversion symmetry of the lattice Hamiltonian in the absence of shaking as noted in Ref.~\cite{ledesma2023machinedesigned}.  Shaking breaks this symmetry, leading to offsets of these regions from the origin, with the sign of the offset depending on the directionality of the initial shaking actions.  We expect such considerations to apply generally when target parameters obey similar symmetries of the Hamiltonian under consideration.   

In Fig.~\ref{fig:2d_fringes}(a) it is clear that there are in fact numerous diagonals and off-diagonals, leading to a visible gridded pattern, which is a signature of \textit{aliasing}.  While visually obvious in our JSD plots,  aliasing limits the accuracy of our sensor, since 
target parameters that fall on different diagonals or off-diagonals can have similar likelihood functions and thus be difficult to distinguish via Bayesian estimation.  Thus, for accurate and efficient determination of acceleration by evaluating a maximum likelihood estimate, one should be careful to restrict Bayesian updating to the parameter domain within distinct cells of the aliasing grid. We expect that such aliasing is a generic feature, present regardless of whether a DSP or SPP is used. 


In addition to information about aliasing, Fig.~\ref{fig:2d_fringes}(a) reveals additional information about the (broader) effective range of the DSP's precision.
This can be seen in the loss of contrast in the aliasing grid that occurs (roughly) for accelerations exceeding $ 0.11g$ in magnitude.  This loss of contrast can be quantified by evaluating $I_a$ locally along a horizontal slice of Fig.~\ref{fig:Jenson-Shannon}(a) as shown in panel (d).  Here the effective range appears as a slowly varying envelope on top of the local classical Fisher information, implying a loss of precision as we move away from $a/g=0$. We see this decay in performance reflected in $D_{JS}(P(p|a,V_L)||P(p|a'=0,V_L)$ which is also shown in panel (d).  Here, the overall envevlope of the JSD function increases as the envelope of $I_a$ decreases.
The decay in $I_a$ is partly due to the DSP being optimized for measuring acceleration's around $a/g=0$ as a phase measurement in the reference frame falling with the atoms (see App.~\ref{app:augmented}, Eq.~\eqref{eq:freefall}). In particular, the values $a/g= \pm 0.11$ coincide with  $2\pi$ phase difference of the phase of the lattice.  Similar considerations will apply to any DSP where the target parameter relies on a phase measurement.

Next, we plot the JSD for fixed acceleration and varying lattice depth in Fig.~\ref{fig:Jenson-Shannon}(b).  
We see that the curvature of the JSD away from the diagonal blue line results in a slow variation in the JSD as a consequence of low $I_{V_L}$, indicating that the likelihood functions in Fig.~\ref{fig:2d_fringes} vary slowly over the considered values of lattice depths. Furthermore, this curvature is a reflection of the slow variation of the measurement outcomes found in Fig.~\ref{fig:2d_fringes} for all momentum as the lattice depth is varied at fixed acceleration. Thus, while the same considerations about aliasing and effective range apply to variations in the lattice depth, they do not affect estimation of acceleration via Bayesian updating.

In order to reveal the decorrelation properties of the DSP, we now evaluate the JSD for simultaneous variation of the lattice depth and acceleration as shown in Fig.~\ref{fig:Jenson-Shannon}(c).  Explicitly, we evaluate the JSD as a function of $P(a,V_L)$ and $P(a',V_L')$ for fixed $a/g = 0$ and $V_L/E_R = 10$.  This form was chosen because the curvature and corresponding CFIM elements determine the values of the JSD around the origin according to Eq.~\eqref{eq:curvature} and the results shown in Fig.~\ref{fig:fi_a}.  Qualitatively, as acceleration is varied at fixed lattice depth, we find many distinct horizontal regions of low-valued JSD.  The narrowness of these regions as a function of $a$ reflects the high sensitivity (large $I_a$) of the DSP to the target parameter.  Similarly, these regions are broad as a function of $V_L$ due to low $I_{V_L}$.  The decorrelating effect of the DSP is visible in an absence of noticeable tilt in the central region due to the small $\mathcal{I}_{a,V_L}$ and dimensionless correlation.  The degree to which each region is tilted is however not uniform as both acceleration or lattice depth is varied away from the point $a/g = 0$, $V_L/E_R = 10$ where the DSP was optimized.  Importantly, this figure provides also additional performance information about the effective range over which the decorrelating effect of the sensor applies.  This information is encoded in the parameter region beyond which the tilt reappears, indicating the development of significant correlations between target and nuisance parameters.

\section{Conclusion}
In this work, we have shown how target signals can be decorrelated from fluctuations of nuisance parameters by desinging sensing protocols using machine learning and multiparameter estimation theory.  This was demonstrated in the specific case of performing accelerometry on an optical lattice interferometer, however the formulation of information-theoretical goals in this work is broadly applicable to decorrelation tasks in quantum metrology and computing.  In addition to generating a concrete example of a decorrelated sensing protocol (DSP), we also perform statistical analysis on the measurement outcomes.  Here visual signatures of the decorrelating effect of the sensing protocol were found as a consequence of the connections between the Fisher information matrix, measures of informational differences between measurements, and curvatures on statistical manifolds.  Additionally, we study the Jensen-Shannon divergence and establish the effective ranges for a DSP for operating as an accelerometer with ensured decorrelation.  A simple scheme for lattice depth estimation is also analyzed in App.~\ref{app:Lattice_JSD} for completeness.

Although DSP's were demonstrated in this work for a specific platform, the method of providing machine learning with knowledge of the information geometry of a quantum system in order to arrive at states which are metrologically advantageous as well as robust is quite general.  In particular, the search for protocols that yield robust gates (c.f. Refs.~\cite{PhysRevResearch.5.033052,jandura2022time,robust_qoc}) and optimal use of quantum resources \cite{kaubruegger2019variational,marciniak2022optimal,PRXQuantum.4.020333} is a highly active topic where these methods may find fruitful applicability.  Additionally, the single target and nuisance parameter example studied in this work can be extended straightforwardly to the many-parameter setting using the general formalism of Sec.~\ref{sec:decsens}.  This setting describes metrological applications in both high-precision metrology as well as the operation of quantum devices in noisy, real-world environments where intense efforts are underway to transition quantum technologies out of the lab and into the field.

\begin{acknowledgments}
The authors acknowledge Catie LeDesma, Kendall Mehling, John Cooper, Chris Wilson, Pranav Gohkale, Maximillian Seifert, Dana Anderson, Zack Pagel, and Noah Fitch for fruitful discussions.   This research was supported in part by NASA under grant number 80NSSC23K1343, NSF OMA 2016244, NSF OSI 2231377, and by a Sponsored Research Agreement between the University of Colorado Boulder and Infleqtion.
\end{acknowledgments}

\bibliography{References}
\begin{appendix}

\section{Augmented states}\label{app:augmented}

In order to evaluate the Fisher information matrices, one must compute inner products involving the derivatives of the quantum state $|\partial_{b^\mu}\psi\rangle$ with respect to the parameters of the model.  This can be done in a variety of ways.  In the absence of a lattice, it can be done analytically as discussed in App.~\ref{app:optqfi}.  In the presence of a lattice, one must resort to numerical methods for instance by running the simulation at parameter values ${\mathbf{b}}$, ${\mathbf{b}} - d{\mathbf{b}}$, and ${\mathbf{b}} + d{\mathbf{b}}$ in order to evaluate the state derivatives via finite differencing.  This is what was done in Ref.~\cite{chih2022train}.  Alternatively, one can form an augmented state composed of the quantum state and the state derivatives $(|\psi\rangle, |\partial_{b^1}\psi\rangle, |\partial_{b^2}\psi\rangle,\dots) $.  Utilizing this method of augmented states (c.f.~\cite{PhysRevApplied.17.014036,PhysRevResearch.5.033052}), one obtains the following equations of motion by taking partial derivatives of the Hamiltonian (Eq.~\eqref{eq:hamiltonian})
\begin{align}
&i\hbar \frac{d}{dt}|\psi\rangle = \hat{H}(t)|\psi\rangle,\label{eq:schro}\\
&i\hbar \frac{d}{dt}|\partial_{b^\mu}\psi\rangle = \hat{H}(t)|\partial_{b^\mu}\psi\rangle + (\partial_{b^\mu} \hat{H}(t))| \psi\rangle.\label{eq:eomparam}
\end{align}
The partial derivative of the Hamiltonian with respect to the lattice depth is given by 
\begin{align}\label{eq:partialV0}
& \partial_{V_L}\hat{H}(t) = -\frac{1}{2}\cos(2k_L \hat{x} + \phi(t)).
\end{align}
To evaluate the partial derivative with respect to the acceleration, we perform the frame transformation discussed in Ref.~\cite{ledesma2023machinedesigned}.  Here, application of the Galilean boost generator $\hat{\mathcal{K}}(t) = \exp(-imat\hat{x}/\hbar)$ yields the transformed Hamiltonian \cite{jordan2012linear}
\begin{align}\label{eq:transfham}
\hat{H}_\mathrm{acc}(t) &= \hat{\mathcal{K}}(t)\hat{H}(t)\hat{\mathcal{K}}(t)^\dagger - i\hbar \left[\partial\hat{\mathcal{K}}(t)/\partial t\right] \hat{\mathcal{K}}(t)^\dagger \nonumber\\
&= \frac{\hat{\pi}^2(t)}{2m} - \frac{V_L}{2} \cos\left[2k_L\hat{x} + \phi(t)\right].
\end{align}
where the inertial force $ma\hat{x}$ is now absent and the kinematic momentum becomes $\hat{\pi}(t) \equiv \hat{p} - mat$.   
In this frame, one finds
\begin{equation}
\partial_a \hat{H}^{(1)}(t) = -\hat{p}t \label{eq:partiala}
\end{equation}
where we ignore a term contributing a constant to the transformed Hamiltonian and note that Eq.~\eqref{eq:partialV0} is unchanged ($ \partial_{V_L}\hat{H}(t) =  \partial_{V_L}\hat{H}_\mathrm{acc}(t)$) in the boosted frame.  Finally, we note that an additional frame transformation was discussed in Ref.~\cite{ledesma2023machinedesigned} where hte acceleration appears instead as an additional phase shift.  Here, the transformation is to the reference frame falling with the atoms, yielding
\begin{equation}\label{eq:freefall}
\hat{H}^{(2)}(t) = \frac{\hat{p}^2}{2m} - \frac{V_L}{2}\cos\left[2k_L + \phi_a(t)\right].
\end{equation}
where $\phi_a(t) = \phi(t) - k_L at^2$.   Thus, measuring acceleration amounts to a phase measurement $\phi_a$ around $a=0g$. This also implies that the $\phi_a$ phase wraps evey $a_0=\frac{2\pi}{k_L t^2}$, treating acceleration values that differ by an $a_0$ shift equally.

While the state vector has unit norm, the norm of the state derivatives, which is initially set to zero, is not fixed due to the source term in Eq.~\eqref{eq:eomparam}, which is responsible for generating quantum Fisher information, related to parameter $b^\mu$ along the quantum trajectory.  Generally, the dynamics of the augmented quantum state vector has a geometrical interpretation as the system moves along a trajectory in Hilbert space and acquires sensitivity to the parameters of the system.  In practice, the augmented state vector can be propagated in time using an RK4 algorithm.  


\section{Free-space Fisher information bounds}\label{app:optqfi}

In this section, we derive the result $I_\mathrm{RI}(p_0) = 4 I_\mathrm{MZI}(p_0)$ for the Fisher information in the Ramsey configuration discussed in Sec.~\ref{sec:refsys} in connection with Ref.~\cite{PhysRevA.98.023629}.  This result can be analytically derived in free space using the augmented state method outlined in App.~\ref{app:augmented}  Eqs.~\eqref{eq:schro} and \eqref{eq:eomparam} can be calculated analytically as we now discuss.

We begin by considering an arbitrary initial state $|\psi(t=0)\rangle$, and then analytically obtain the evolution of the associated augmented state vector by solving Eqs.~\eqref{eq:schro} and \eqref{eq:eomparam}. The Fisher information at the terminal time $\mathcal{T}$ can be then be straightforwardly extracted.  The equation of motion for the quantum state is given by Eq.~\eqref{eq:schro}, which can be solved as $|\psi(t)\rangle = \hat{U}(t)|\psi(t=0)\rangle$, where $\hat{U}(t) = \exp(-i\int^t ds \hat{H}(s))$ is the evolution operator
\begin{align}
    \hat{U}(t) = & \exp\left[-\frac{i}{2m\hbar}\int^{\mathcal{T}}_{0} ds\ \hat{\pi}(t)^2\right],\\
    =& \exp{\left[-\frac{i}{2m \hbar}\left(\hat{p}^2\mathcal{T}-ma\mathcal{T}^2\hat{p} + \frac{m^2a^2\mathcal{T}^3}{3}\right) \right].}
\end{align}
Likewise, the equation of motion (Eq.~\eqref{eq:eomparam}) for the state derivative $|\partial_a \psi(t)\rangle$ can be solved as
\begin{align}
|\partial_a\psi(t)\rangle = &\ \partial_a \hat{U}(t) |\psi(0)\rangle,\nonumber\\ 
=& -i \hat{\mathcal{G}}_a(t) \hat{U}(t)|\psi(0)\rangle
\end{align}
where 
\begin{align}
\hat{\mathcal{G}}_a(t) =& \int_0^t ds\ \partial_a\hat{H}(s),\nonumber\\
= &-\frac{t^2}{2\hbar}\left(\hat{p}-\frac{2 ma t}{3}\right)\label{eq:Ggenerator}
\end{align}
is the generator associated with the acceleration \cite{PhysRevA.98.023629}. Note that this generator is the same in origin as the source term in the App.~\ref{app:augmented}.

At time $t=\mathcal{T}$, one obtains
\begin{equation}\label{eq:qmstateaug}
    \ket{\partial_a\psi(\mathcal{T})}= \left[-\frac{i}{2m\hbar} \left(\frac{2am^2\mathcal{T}^3}{3}-m\mathcal{T}^2\hat{p}\right)\right]\ket{\psi(\mathcal{T})},
\end{equation}
We see from the generator that acceleration encodes a phase difference between two momentum states based on their momentum difference. Thus, a maximal phase difference can be generated by choosing an equal superposition of two momentum states with a high momentum splitting.

To obtain the Fisher information, we specify an initial condition $\ket{\psi(t=0)} = \ket{\pm p_0}$ for the instantaneous application of an ideal beam-splitter to the the quantum state at $t=0$.  In this state, $\langle \hat{p}\rangle = 0$ and $\langle \hat{p}^2\rangle = p_0^2$, and Eq.~\eqref{eq:qfimoffdiag} gives $
    F_\mathrm{RI}(p_0)= \left(\frac{p_0\mathcal{T}^2}{\hbar}\right)^2
$.  The instantaneous application of an ideal recombiner at $t=\mathcal{T}$ achieves $I_\mathrm{RI}(p_0) = F_\mathrm{RI}(p_0)$ \cite{PhysRevA.98.023629}.

\section{Reinforcement learning}\label{app:rl}

Reinforcement learning is a particularly well suited for method for generating control functions for quantum problems, such as the end-to-end construction of DSP's considered in this work, where the dimensionality of the problem can become very high and the memory footprint of quantum optimal control methods becomes prohibitive~\cite{chih2022train, PhysRevResearch.3.033279, ledesma2023machinedesigned, 10015539, 9867736, 10156455,PhysRevLett.120.263201, Weidner_2018}.  We employ a specific, model-free kind of reinforcement learning using the double-deep-Q approach \cite{sutton2018reinforcement, Hasselt_Guez_Silver_2016}, which has been applied previously to design control functions for optical lattice interferometers \cite{chih2022train, PhysRevResearch.3.033279, ledesma2023machinedesigned}.  Briefly, the reinforcement learning agent selects an action $\mathbf{a}$ when exposed to the state $\mathbf{s}$ of an environment. The environment subsequently carries out that action and returns to the agent a reward $\mathbf{r(s')}$ based on the ``quality'' of the resulting state $\mathbf{s'}$.  We use Q-learning for the decision making process of the agent which involves the evaluation of quality factors, called Q-values $\mathbf{Q(s,a)}\in\mathbb{R}$, using neural networks, as depicted Fig.~\ref{fig:conceptual}, where each element represents the desirability of a possible action $\mathbf{a}$ given a state $\mathbf{s}$. For action selection, the agent uses a policy $\pi$ based on the Q-value to decide the next action - for our case, we employ a greedy policy selects the action corresponding to the largest Q-value for the current state.  To strike a balance between learning and exploration, we employ epsilon greedy exploration, where the agent decides the next action either randomly or based on $\pi$ using a probability $\epsilon$. The probability of exploration $\epsilon$ is steadily decreased with subsequent training episodes.

As discussed in Sec.~\ref{sec:decsens}, the goal of a DSP is to optimize sensitivity to a target parameter while minimizing the effect of nuisance parameters.  Therefore, we formulate rewards as $\mathbf{r}[\mathcal{I}]$.  We comment that Ref.~\cite{chih2022train}, considered the reward $\mathbf{r(s')} = I_{b^1}$, appropriate for single-parameter estimation with perfect information about all other parameters of the system, producing control functions maximizing single-parameter sensitivity (SPP's).  For DSP's, the agent requires access to {\it more} information in order to perform the dual tasks of sensitivity optimization {\it and} decorrelation.  Here, the natural candidate is to construct reward functions for DSP's from elements of Eq.~\eqref{eq:BlockInversion}, which ultimately determine the theoretical sensitivity of the device in the multiparameter setting (see Eq.~\eqref{eq:ccrb}).  Explicitly, the results of Secs.~\ref{sec:mpqs} and \ref{sec:stats} were produced using $\mathbf{r}[\mathcal{I}] = f[\mathcal{I}]/(1-f[\mathcal{I}])$ with $f[\mathcal{I}]=(2 I_\mathrm{MZI}(\mathcal{I})^{-1}_{a,a})^{-1} = \zeta_a/2$.  We note that this can be translated to a reward function for SPP's as the assuption of perfect information about untargeted parameters produces the mapping $\mathcal{I}^{-1}_{a,a}\to  1/I_a$  Additionally, DSP-appropriate cost functions for quantum control methods can be constructed analogously although we do not pursue that avenue in this work.  
In Q-learning, the necessary condition that must be satisfied in order to describe optimality is given by the Bellman equation \cite{sutton2018reinforcement}
\begin{equation}\label{eq:bellman}
\mathbf{Q_\pi(s,a)} = \mathbf{r(s')} + \mathbf{\gamma\ \underset{a'}{\mathrm{max}}\ Q_\pi(s',a')}.
\end{equation}
This equation states that optimality is achieved when the Q-value ($\mathbf{Q(s,a)}$) of taking action $\mathbf{a}$ in state $\mathbf{s}$ under policy $\mathbf{\pi}$ is equal to the Q-value ($\mathbf{Q(s',a')}$) in the expected next state $s'$ discounted by the factor $\mathbf{\gamma}$ and maximized over all possible actions $\mathbf{a'}$ under $\mathbf{\pi}$ plus the reward received ``along the way" $\mathbf{r(s')}$.  Essentially, the value of taking a particular action in a particular state is determined by the reward received in the following state plus the value of making optimal decisions informed by the Q-values thereafter.  

The Bellman equation is a recursive relation between Q-values and in low-dimensional problems, a tabular approach can be taken to record and iterate the Q-values \cite{sutton2018reinforcement}.  However due to the enormous dimensionality of the present problem, we employ neural networks in the double-deep-Q-approach and train using repeated runs of a numerical simulation or experiment.  In practice, two neural nets are used in order to evaluate Eq.~\eqref{eq:bellman} and provide additional stability \cite{Hasselt_Guez_Silver_2016, mnih2015human,PhysRevResearch.3.033279}, the Q-network and a target network.  The Q-network is used to evaluate $\mathbf{Q(s,a)}$ while $\mathbf{Q(s',a')}$ is evaluated using the target network, and then the difference between the left and right hand sides of the Bellman Equation (Eq.~\eqref{eq:bellman}) is evaluated and backpropagated through to update the Q-network.  The target network is then updated via a weighted average, with weighting $\mathbf{\tau}$.  Additional details of the double-deep-Q algorithm applied to optical lattice interferometry can be found in Refs.~\cite{PhysRevResearch.3.033279, chih2022train}. 

As indicated in Table~\ref{tab:hyperparameters}, each neural net has a single hidden layer with 64 nodes all fully connected.  The set of actions corresponds to 16 possible discrete amplitude values
\begin{equation}\label{eq:actions}
\mathbf{\mathcal{A}} = \{n\pi / 12| n = 0, 1,\dots, 15\}.
\end{equation}
These actions determine the control parameters $\mathbf{c}$ which translate into shaking over a half period oscillation of the form $\phi(t) = \mathrm{Amp}(t)\times \sin(\omega_s t)$ where $\omega_s = 11.5\omega_R$ with $\omega_R = E_R/\hbar$.  The time interval between transitions is then $\pi/11.5\approx 0.27 \omega_R$.  This is a variant of the action set considered in Ref.~\cite{PhysRevResearch.3.033279} where it was found to work well in driving transitions between low-lying Bloch states.

The selection of features $s$ to input in the neural net requires physical intuition.  In Ref.~\cite{PhysRevResearch.3.033279}, the momentum-space occupations were used.  In Ref.~\cite{chih2022train}, position-space expectation values to provide the agent additional information about angular momentum for rotational sensing.  From Eq.~\eqref{eq:hamiltonian}, we see then that the expectation value $\langle \hat{x}\rangle$ should be added to the feature set.  Physically, this expectation value provides information to the agent about the free-fall of the non-inertial frame as appropriate for accerometry.  In App.~\ref{app:augmented}, it was shown how the acceleration enters the Hamiltonian proportional to $t$ (see Eq.~\eqref{eq:transfham}).  The features consist then of the dimensionless flag $t/\mathcal{T}$ along with the parity-sensitive populations of states $|p,\pm\rangle$, which provides additional information allowing the agent to distinguish, for instance, between nearly-degenerate, opposite-parity Bloch states in the valence band.  The flag $t/\mathcal{T}$ also allows the agent to distinguish between mirrored matter-wave trajectories, for instance in the Mach-Zehnder configuration.

\begin{table}
\caption{\label{tab:hyperparameters}Choice of hyperparameters for the design of DSP's. The exploration probability decays exponentially for each episode between 1.0 and 0.1 at the
rate represented by $\mathbf{\epsilon}$ decay. The ``Adam'' optimizer is a commonly-used method for
stochastic gradient descent \cite{kinga2015method}.}
\begin{ruledtabular}
\begin{tabular}{lr}
    \toprule
    Hyperparameters & Values \vspace{1mm}\\
    \hline
    \vspace{-3mm}\\
    $\mathbf{\gamma}$ & 0.99\\
    $\mathbf{\tau}$ & 0.8\\
    $\mathbf{\alpha}$ & $10^{-3}$\\
    episodes & 5000\\
    $\mathbf{\epsilon}$  decay & $2.5\times 10^{-5}$\\
    Hidden size & 64\\
    Batch size & 100\\
    Optimizer & Adam\\
    \bottomrule
\end{tabular}
\end{ruledtabular}
\end{table}

\section{Expansions of the Jensen-Shannon divergence}\label{app:jsdexpand}

We provide further details on the analytic derivation of the expansions of the Jensen-Shannon divergence and connections with the CFIM given in Sec.~\ref{sec:stats}.  We begin by expanding $D_{JS}(P({\bf b})||P({\bf b} + d{\bf b}))\equiv f({\bf b} +d{\bf b}) $ in a Taylor series to second order in the differential parameter vector $d{\bf b} = (db^1, db^2,...)$, yielding
\begin{equation}\label{eq:taylor}
f({\bf b}+d{\bf b})\approx  f({\bf b}) + [\nabla_\mu f({\bf b})] db^\mu + \frac{1}{2} H_f({\bf b})_{\mu\nu} \ db^\mu db^\nu,
\end{equation}
where $H_f({\bf b})$ is the (symmetric) Hessian matrix whose elements are defined as
$(H_f({\bf b}))_{ij} \equiv \partial^2 f/\partial b^i\partial b^j$.

We discuss the evaluation of each term in the expansion in Eq.~\eqref{eq:taylor}.  The contribution $f({\bf b})$ is vanishing by virtue of the definition of the JSD (Eq.~\eqref{eq:jsd}).  Likewise, the contribution of the gradient 
also vanishes because the minimum of the JSD is absolute.  The Hessian matrix is however nonzero, making the contribution
\begin{equation}
H_f({\bf b})_{\mu\nu} = \frac{\mathcal{I}_{\mu\nu}}{4}.
\end{equation}
which establishes the connection between the CFIM and the local curvature of the JSD, quantified by the Fisher information metric (Eq.~\eqref{eq:scorecfim})
\begin{equation}
D_{JS}(P({\bf b})||P({\bf b} + d{\bf b}))\ \approx \frac{1}{8} \mathcal{I}_{\mu\nu} \ db^\mu db^\nu \Big|_{\bf{b}}.
\end{equation}
This metric can be used to calculate information distances on smooth statistical manifolds \cite{cover1999elements}.  In the case considered in this work, only two components of the differential element $d{\bf b} = (da,dV_L)$ are varied and one finds Eq.~\eqref{eq:curvature}.


\section{Decorrelated sensing protocol for lattice depth estimation}
\label{app:Lattice_JSD}

\begin{figure}[t!]
	\centering
    \includegraphics[clip, width = 1 \columnwidth]{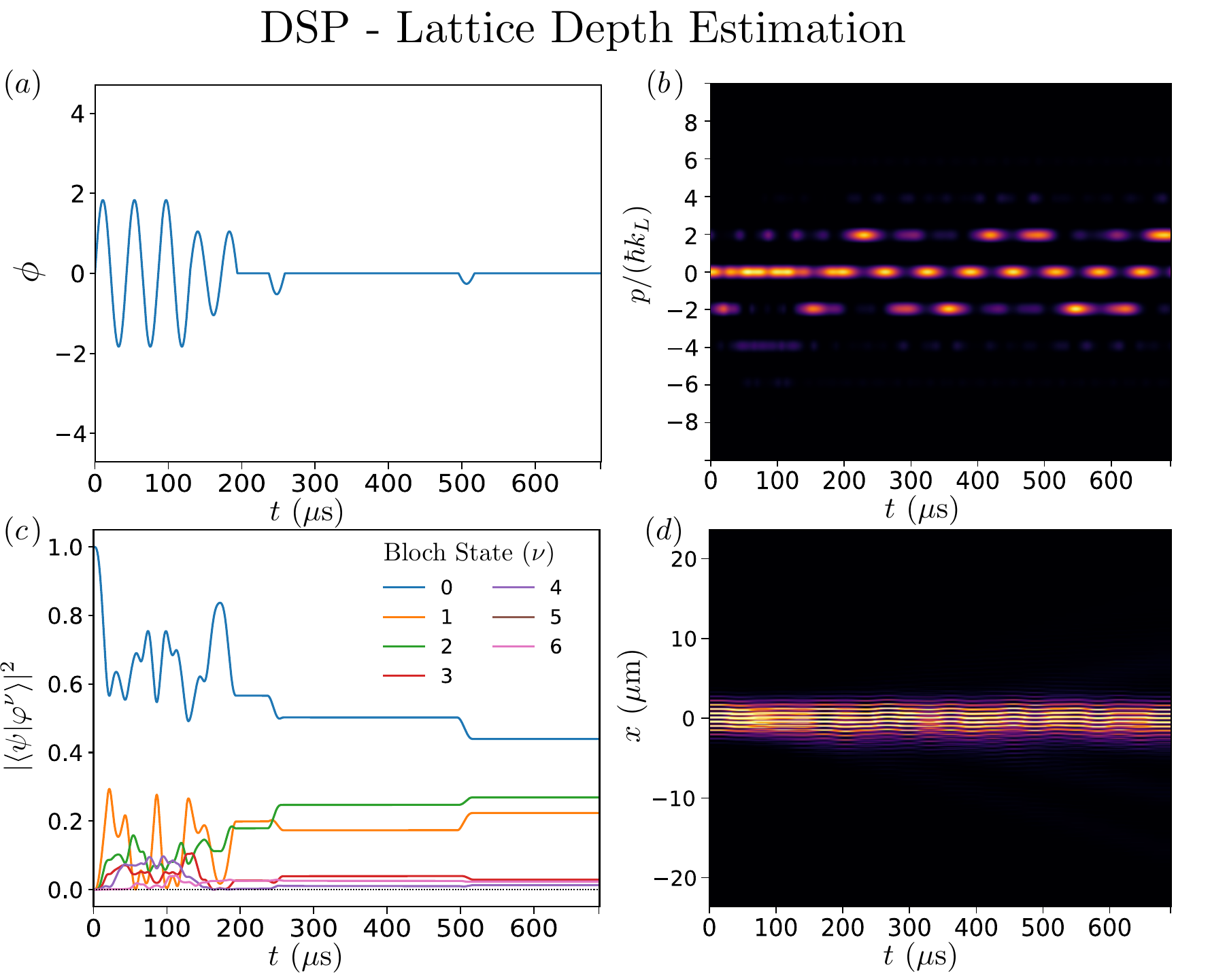}
    \caption{Lattice depth estimation using a DSP, including the control function (a) momentum space dynamics (b), Bloch state occupations (c), and position space dynamics (d).  The position space visualizations use the same convolution method described in the caption of Fig.~\ref{fig:evo_a}.  In practice the wave packet in an optical lattice interferometer is delocalized over many lattice sites.}		\label{fig:evo_V0}
\end{figure}

\begin{figure}[t!]
	\centering
    \includegraphics[clip, width = 1 \columnwidth]{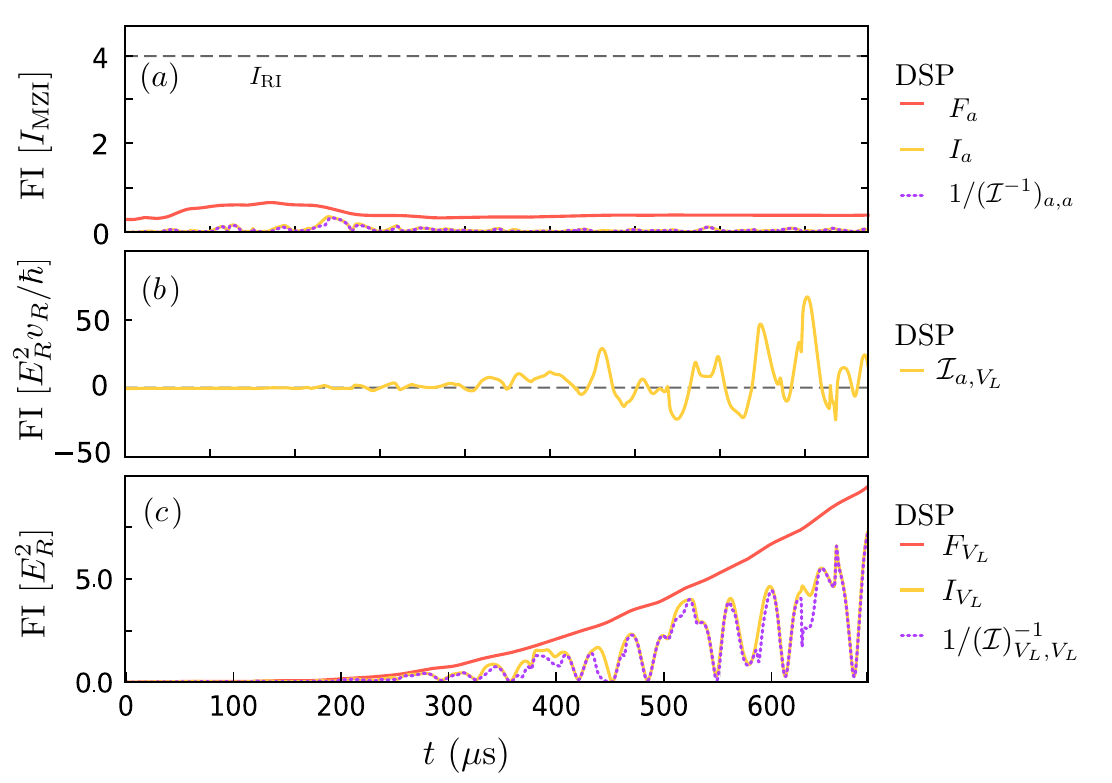}
    \caption{Relevant classical and quantum Fisher information (FI) elements for (a) acceleration, (b) covariance, and (c) lattice depth for the DSP designed for lattice depth estimation.  In (c) a $F_{V_L}\propto t^3$ power is found once the valence band superposition is prepared (see Fig.~\ref{fig:evo_V0}(c)).  Although not aimed at accelerometry, we provide in panel (a) the same rescaling in units of the Mach-Zehnder configuration result with dashed line indicating the optimal result in the Ramsey configuration discussed in Sec.~\ref{sec:refsys}.} \label{fig:fi_V0}
\end{figure}


Sections~\ref{sec:mpqs} and \ref{sec:stats} presented and analyzed a DSP for accelerometry with lattice depth treated as a nuisance parameter.  It is however valuable to have a precise estimate of $V_L$ prior to performing quantum sensing or simulation tasks \cite{doi:10.1126/science.aal3837,RevModPhys.86.153,RevModPhys.80.885,bloch2012quantum}.  Here, the strength $a$ of the inertial force can be viewed instead as a nuisance parameter as the device may be operating in an un-characterized or fluctuating inertial environment.  In this section, we present and analyze a DSP for this scenario, placing the result in the context of existing methods, which typically provide uncertainties in the $10 -20 \%$ range \cite{PhysRevA.57.R20,PhysRevLett.83.284,PhysRevLett.56.827,PhysRevA.65.063612}.  

The same set of actions (see App.~\ref{app:rl}) are available to the agent for lattice depth estimation as for accelerometry, and the reward function $\mathbf{r}[\mathcal{I}]$ is again a function of the CFIM.  However, we use an alternative non-linear function to boost large rewards and speedup learning $\mathbf{r}[\mathcal{I}] = \sinh(4 / (\mathcal{I})^{-1}_{V_L,V_L})$.  We see in Fig.~\ref{fig:evo_V0}(a) that the found DSP is largely zero after an initial repeated pattern.  We emphasize that a succession of repeated values found by the agent is a highly-nontrivial pattern as the Q-values that the agent uses for decision making are randomly chosen from the $16^{32}$ possible distinct control functions.  The initial forays of the agent in a reinforcement learning algorithm are also exploratory in nature before becoming less random as the training iterates over many episodes (see App.~\ref{app:rl} and \cite{sutton2018reinforcement}).  

In Fig.~\ref{fig:evo_V0}(b-d), we see that the DSP yields a {\it periodic} evolution in momentum space after the quantum state is placed in a superposition of valence bands.   From Fig.~\ref{fig:evo_V0}(d) it is clear that transport of the wave packet between sites is suppressed, which results from the gapped nature and weak dispersion of the valence bands.  This periodic oscillation between momentum components is heavily reminiscent of the dipole oscillations studied in Ref.~\cite{PhysRevA.97.043617}, and we note that center of mass displacements can be seen even in Fig.~\ref{fig:evo_V0}(d).    Consequently, decorrelation is added to the list of robust features of this method.

We now perform a Fisher analysis to understand why this method is favored as well as how it achieves decorrelation.
Figure~\ref{fig:fi_V0} shows the Fisher information matrices where it is clear that this DSP performs poorly as an accelerometer.  On the other hand, we find monotonic growth in the lattice-depth sensitivity $F_{V_L}\propto t^3$ once the state is prepared in the valence band superposition.  Here, the actions serve to time a maximum of $I_{V_L}$ with measurement, aided by the two nonzero actions following the initial pattern.  

In terms of decorrelation, the DSP minimizes the penalty due to the nuisance parameter $a$, with $\mathrm{Corr}(\ell_a, \ell_{V_L})\approx 0.081$.  How does the sensitivity of this DSP compare to traditional methods?  Quantitatively, the result $I_{V_L}E_R^2=7.31$ yields an uncertainty in the lattice depth of $0.37 E_R$, providing a lower bound on the single-shot error of $\sim 3\%$ for lattice depth estimation.  This is comparable to the experimental results of Ref.~\cite{PhysRevA.97.043617} using the sudden phase shift method, however in that case many shots were required to achieve the fringe visibility required for lattice depth estimation.

\begin{figure}[t!]
    \centering
    \includegraphics[width=\linewidth]{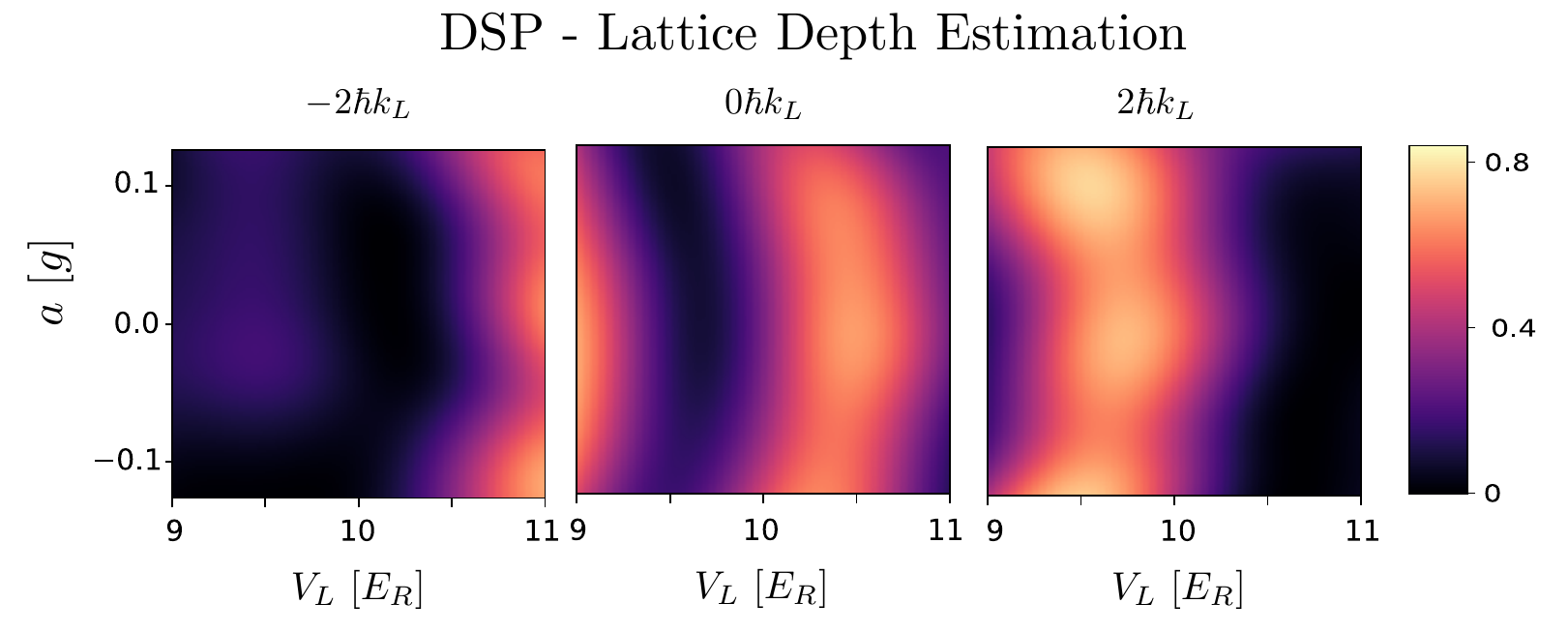}
    \caption{Final momentum probability distributions $P(p|a,V_L)$, resulting from the DSP for lattice-depth estimation, and presented over a range of $a$, $V_L$, and $p$.  We display occupations up to momenta $\pm 2\hbar k_L$ and note that for larger momenta are not significantly occupied in this DSP as the lattice depth or acceleration are varied significantly.}
    \label{fig:2d_fringes_lattice}
\end{figure}

We now provide a brief statistical analysis of these results, extending the discussion of Sec.~\ref{sec:stats} to lattice depth estimation.  The momentum probability distributions $P(p|a,V_L)$ are given in Fig.~\ref{fig:2d_fringes_lattice}.  Comparing to Fig.~\ref{fig:2d_fringes}, we note the reduced contrast and periodicity as acceleration is varied at fixed lattice depth.  This is a reflection of the poor performance of the lattice-depth DSP as an accelerometer--as expected.  However, The variation of the distributions with lattice depth at fixed acceleration is more dramatic--again due to the different design goals.  Consequently, care should be taken when Bayesian inferencing for lattice depth estimation to restrict to smaller regions of $V_L$.   

\begin{figure}[t!]
    \centering
\includegraphics[width=\linewidth]{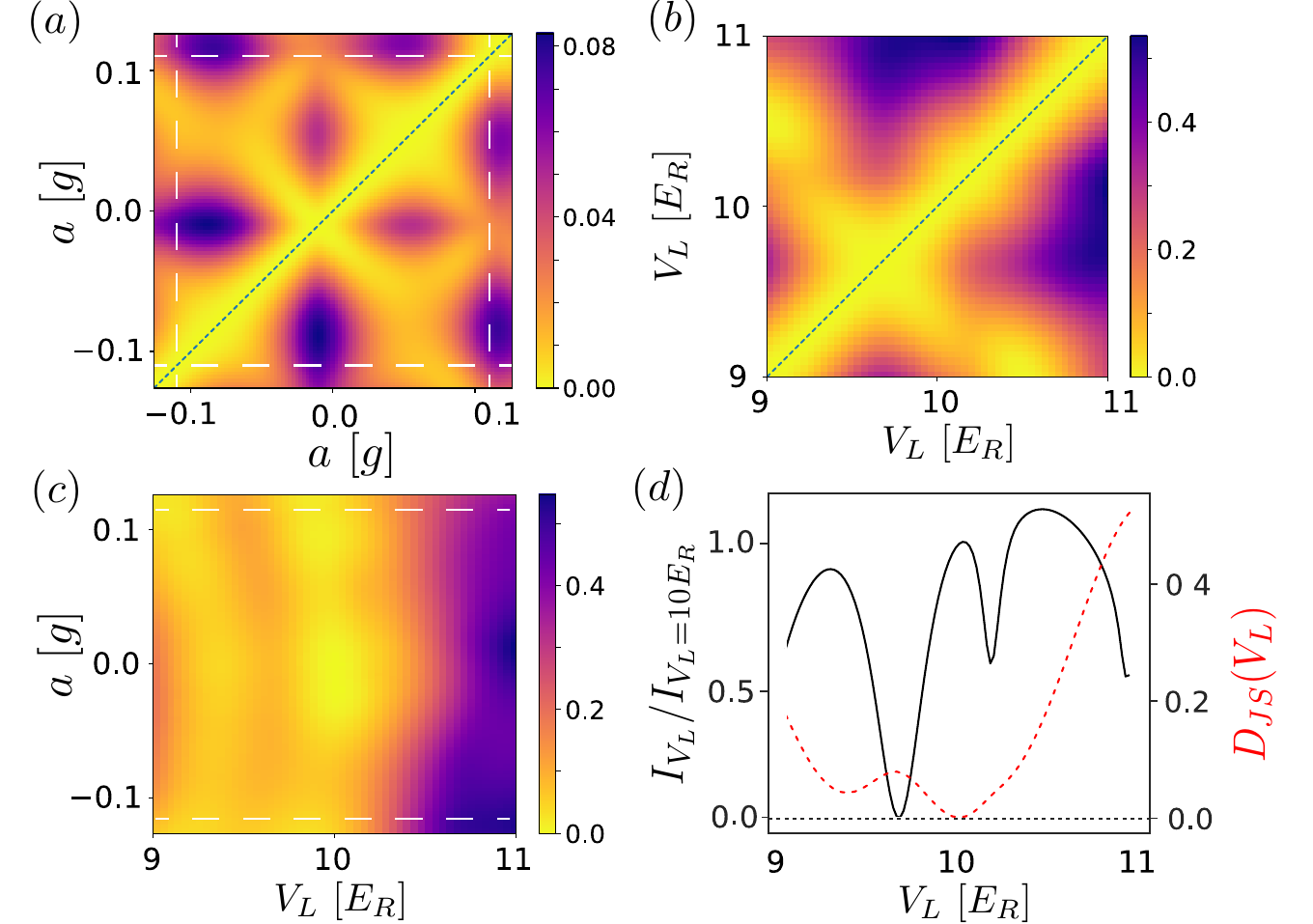}
    \caption{Visualization of the Jensen-Shannon divergence for the lattice-depth focused DSP as pairs of parameters are varied, including (a) fixed lattice depth and varying acceleration, (b) fixed acceleration and varying lattice depth, and (c) varying acceleration and lattice depth.  As in Fig.~\ref{fig:Jenson-Shannon}, the non-varied parameters are held fixed at $a/g = 0$ and $V_L/E_R=10$ although a different DSP is used in this case.  In panels (a) and (c), dashed white lines indicate the effective range of acceleration for the DSP.  In (b) a larger portion of the aliasing grid is shown for the lattice depth however the effective range for the DSP is not reached over the parameter range considered.  In (c) the cross indicates the origin whose local curvature matches the CFIM elements presented in Fig.~\ref{fig:fi_V0}.  The green dotted lines in (a) and (b) indicate the main diagonals where the JSD vanishes.  (d) Evaluation of the JSD (dotted red line) as $D_{JS}(V_L||V_L')$ with $V'_L/E_R=10$ and fixed acceleration, along with the local $I_{V_L}$ evaluated along the $V_L = 10E_R$ horizontal of panel (b). The effective range of the DSP for performing lattice depth estimation is not exceeded over the range of $V_L$ shown, such that an envelope analogous to the one found in Fig.~\ref{fig:Jenson-Shannon}(d) is not yet visible.}
    \label{fig:JSD_lattice}
\end{figure}

Furthermore, we compress the multi-fringed probability distributions into entropy measures using the Jenson-Shannon divergence with results shown in Fig.~\ref{fig:JSD_lattice}.  First, we note that in panel (a) the aliasing grid pattern is harder to observe compared to the accelerometry DSP.  In particular, diagonal lines where the JSD achieves low values are significantly broader now as a consequence of the lower values of $I_a$ compared to Fig.~\ref{fig:Jenson-Shannon}(a).  The periodicity of the aliasing pattern found in Fig.~\ref{fig:Jenson-Shannon} for accelerometry is here greatly enlarged with only four unit cells visible over the same range presented in Fig.~\ref{fig:Jenson-Shannon}(a). Instead, the aliasing pattern in lattice depth is much more visible as shown in panel (b).  The JSD curvature is also shown in Fig. (c), reflecting the reduced correlation produced by the DSP, larger lattice-depth curvature due to enhanced $I_{V_L}$, and lower acceleration curvature due to reduced $I_a$--as expected.  In panel (d) we show the local Fisher information $I_{V_L}$ as a function of lattice depth, along with $D_{JS}(P(p|a,V_L||P(p|a,V'_L)$ with $V'_L/E_R=10$ and $a = 0g$.  Here, greater lattice-depth sensitivites are found away from the point $(0g, 10E_R)$ at which the machine learning agent was trained. However, decorrelation of the two parameters is still only guaranteed near the training point. Because we do not consider a wide enough range of $V_L$, we do not reach values at which the effective operating range of the sensor in terms of the lattice depth is reached (unlike the remarks made surrounding Fig.~\ref{fig:Jenson-Shannon}(d)).  Here we caution that for larger lattice depths, the competition of interactions and kinetic energy can drive a cold-atom system across an insulating phase transition, which is beyond the scope of the present work \cite{greiner2002quantum}. 
\end{appendix}

\end{document}